\documentclass[12 pt]{article}
\usepackage{bbm}
\usepackage{natbib}
\usepackage{amssymb}
\usepackage{amsmath}
\usepackage{amsthm}
\usepackage[onehalfspacing]{setspace}
\usepackage[colorlinks]{hyperref}
\usepackage{graphicx}
\usepackage{subcaption}
\usepackage{enumitem}
\usepackage{verbatim}
\usepackage{sectsty}
\usepackage[svgnames]{xcolor}
\usepackage{subfiles}
\usepackage[capitalize,nameinlink]{cleveref}
\usepackage[colorinlistoftodos,textsize=tiny]{todonotes}
\usepackage{xcolor}
\usepackage{pgf}
\usepackage{tikz}
\usepackage{xargs}
\usepackage[font={footnotesize,it},justification=justified]{caption}
\usepackage[normalem]{ulem}

\hypersetup{
    pdftitle={The Political Economy of Zero-Sum Thinking},
    pdfauthor={Nageeb Ali, Max Mihm, Lucas Siga},
    citecolor=DarkBlue,
    bookmarksnumbered=true,
    urlcolor=Indigo,linkcolor=DarkBlue
}
 \theoremstyle{definition}
 
 \theoremstyle{remark}

\theoremstyle{plain}
\newtheorem{theorem}{Theorem}

\newtheorem{assumption}{Assumption}

\newtheorem{lemma}{Lemma}

\newtheorem{proposition}{Proposition}

\theoremstyle{definition}
\newtheorem{definition}{Definition}

\renewcommand{\epsilon}{\varepsilon}
\usepackage[margin=1.2in]{geometry}
\newcommandx{\ali}[2][1=]{\todo[linecolor=blue,backgroundcolor=blue!25,bordercolor=blue,#1]{#2}}
\newcommandx{\siga}[2][1=]{\todo[linecolor=red,backgroundcolor=red!25,bordercolor=red,#1]{#2}}  
\newcommandx{\mihm}[2][1=]{\todo[linecolor=green,backgroundcolor=green!25,bordercolor=green,#1]{#2}}

\newcommand{\pop}{\mathcal N}

\newcommand{\inforeal}{\mathcal M}

\newcommand{\trule}{\tau}
\newcommand{\Kappa}{\mathcal{K}}
\newcommand{\vw}{v_w}
\newcommand{\vl}{v_\ell}

\renewcommand{\equiv}{:=}
\renewcommand{\Re}{\mathbb{R}}

\newcommand{\ceil}[1]{\lceil #1 \rceil}
\renewcommand{\succeq}{\succcurlyeq} 

\newcommand{\pg}{p^* }
\newcommand{\pb}{p_* }

\usetikzlibrary{shapes,arrows,automata,fit}

\tikzstyle{info}=[circle,thick,draw=black,fill=black!25,minimum size=4mm]
\tikzstyle{uninfo}=[circle,thick,draw=black,fill=white,minimum size=4mm]
\tikzstyle{inforecog}=[circle,line width=1mm,draw=black!50,fill=black!25,minimum size=4mm]
\tikzstyle{uninforecog}=[circle,line width=1mm,draw=black!50,fill=white,minimum size=4mm]

\tikzstyle{traded}=[draw, line width=1mm]
\tikzstyle{recog}=[draw=black!50, line width=1mm]

\sectionfont{\color{DarkRed}}
\subsectionfont{\color{DarkRed}}
\subsubsectionfont{\color{DarkRed}}
\crefname{assumption}{assumption}{assumptions}
\newcommand{\appendixref}[1]{\hyperref[#1]{Appendix}}
\newcommand{\suppappendixref}[1]{\hyperref[#1]{Supplementary Appendix}}

\allowdisplaybreaks  

\usepackage{nicefrac, xfrac}
\begin{document}

\begin{titlepage}

\title{The Political Economy of Zero-Sum Thinking\thanks{This paper has benefited from the constructive feedback of the Co-Editor and anonymous referees. We thank Peter Buisseret, Wioletta Dziuda, Matt Gentzkow, Ben Golub, Yingni Guo, Jacopo Perego, Philipp Strack, Sevgi Yuksel, numerous seminar and conference audiences, and owe special thanks to Navin Kartik and Richard Van Weelden for their insightful discussions. }}
\author{
S. Nageeb Ali\thanks{Department of Economics, Penn State. Email: \href{mailto:nageeb@psu.edu}{nageeb@psu.edu}.}
\and
Maximilian Mihm\thanks{Division of Social Science, NYU-Abu Dhabi. Email: \href{mailto:max.mihm@nyu.edu}{max.mihm@nyu.edu}.}
\and 
Lucas Siga\thanks{Department of Economics, University of Essex. Email: \href{mailto:lucas.siga@essex.ac.uk}{lucas.siga@essex.ac.uk}.}
}

\maketitle

\begin{abstract}

This paper offers a strategic rationale for zero-sum thinking in elections. 
We show that asymmetric information and distributional considerations  together make voters wary of policies supported by others. This force impels a majority of voters to support policies contrary to their preferences and information. Our analysis identifies and interprets a form of ``adverse correlation'' that is necessary and sufficient for zero-sum thinking to prevail in equilibrium.

\end{abstract}
\thispagestyle{empty} 

\end{titlepage}

\setcounter{tocdepth}{1}

\thispagestyle{empty}
\newpage
\setcounter{page}{1}

\section{Introduction}\label{Section-Introduction}
Zero-sum thinking approaches policy decisions with the mindset that gains that accrue to some parties necessarily come at the expense of others. 
A remarkable feature of this worldview is that it manifests even in a non zero-sum choice, when one policy is widely regarded to benefit voters on average.
For instance, experts see the immigration of skilled labor as spurring economic growth and innovation, creating a larger tax base, and even leading to greater employment of those native-born through complementarities and agglomeration effects.\footnote{According to a poll run by Chicago Booth's Kent Clark Center, $81\%$ of economists surveyed disagreed with the claim that reducing H-1B visas would increase US tax revenues, and $64\%$ disagreed with the claim that it would increase employment for American workers; see \href{https://www.kentclarkcenter.org/surveys/high-skilled-immigrant-visas/}{https://www.kentclarkcenter.org/surveys/high-skilled-immigrant-visas/}.} Nevertheless, a common refrain against immigration stresses that every job taken by an immigrant could otherwise have gone to a native. In trade policy, economists typically emphasize that free trade is mutually beneficial, allowing consumers to obtain goods and services at lower prices \citep[e.g.,][]{SapienzaZingales}. Yet, protectionist arguments focus on how elites within a country gain at the expense of workers. 

This ``us vs. them'' mindset is prominent in populist movements today. Leading up to the Brexit referendum of 2016, the  majority of analysts viewed EU membership as a ``positive-sum'' policy and cautioned that leaving would be detrimental to UK workers, consumers, and businesses.\footnote{These warnings have borne out. In 2024, the Cambridge Econometrics Report commissioned by London City Hall estimated a loss of nearly two million jobs overall in the UK; see \href{https://tinyurl.com/ya8bcyxr}{https://tinyurl.com/ya8bcyxr}. } However, the Leave campaign pointed to the elite's resistance to the reform as evidence that the average citizen would benefit from Brexit. The leader of the UK Independence Party, Nigel Farage, framed the debate as ``a coalition of the politicians against the people'' and declared Brexit to be ``a victory for real people, a victory for ordinary people, a victory for decent people\ldots.''\footnote{See \href{https://tinyurl.com/3pahmkm3}{https://tinyurl.com/3pahmkm3} and \href{https://tinyurl.com/46tnsvhr}{https://tinyurl.com/46tnsvhr}.} The campaign thus exploited a ``zero-sum'' perspective on EU membership to generate support for the reform. Analogously, in his 2016 presidential campaign, the then Republican presidential nominee Donald J. Trump promised to pull the US out of the Trans-Pacific Partnership, which he asserted was ``another disaster done and pushed by special interests\ldots''. He fulfilled his promise once he became president and viewed his executive order as a ``Great thing for the American worker\ldots'' Yet, analyses of trade suggest that the TPP was ``positive-sum'' economically and politically \citep{petri2016economic}.\footnote{More broadly, pertaining to trade, most estimates show that the US-China trade war that began in 2018 has been costly for the average US household \citep{fajgelbaum2022economic}. \cite{autor2024help} find that, contrary to the zero-sum perspective, this trade war did not actually increase employment of American workers, even in the places where the newly-protected industries were located; however, it strengthened political support for the Republican party, particularly in those areas.}

All of the above suggests a puzzle: why does zero-sum thinking prevail even in ``positive-sum''  settings in which voters' interests are largely aligned? 
Prior work \citep[e.g.,][]{meegan2010zero,chinoy2022zero} has largely viewed zero-sum thinking as a psychological trait. In contrast, we offer an informational microfoundation that shows how zero-sum thinking can manifest even with completely rational voters.
%

We study policy choices in which voters do not know how many and who benefits from one policy versus another. We find that distributional uncertainty coupled with asymmetric information gives rise to zero-sum thinking: a voter votes against policies that command the support of others out of fear of what their support means for her. We show that this effect can push an electorate overall to support policies contrary to its interests. This mechanism is purely strategic, emerging independently of identity politics or partisan interests, and even with voters who are ex ante symmetric.

To see how the strategic mechanism works, consider the following example.
Ann, Bella, and Carol vote between policies $p^*$ and $p_*$, and the policy that obtains more votes wins. Each voter's payoff from policy $p_*$ is $0$. By contrast, policy $p^*$ results in two \emph{winners}, each of whom accrue a payoff of $2$, and one \emph{loser}, who obtains a payoff of $-3$. Ex ante, voters are symmetric: there are three equally likely states of the world, $\{\omega_A,\omega_B,\omega_C\}$, where $\omega_i$  is the state in which voter $i$ is the loser. 

Given these payoffs, policy $p^*$ would win the election were voters \emph{fully informed}, i.e., to know the state of the world. Being ex ante optimal, that policy would also prevail if all voters were known to be uninformed, i.e., the \emph{no-information} benchmark.\footnote{Throughout our analysis, we focus on weakly undominated equilibria.} 

Our interest is in the setting in which voters may be asymmetrically informed. Suppose that each voter privately learns her payoff from policy $p^*$ with (independent) probability $\lambda>0$ and otherwise remains uninformed. To illustrate the scope for zero-sum thinking, we envision that $\lambda$ is small, which implies that, with high likelihood, all three voters view policy $p^*$ to be superior. Nevertheless, each voter's uncertainty about what the other voters have learned results in a strict equilibrium that selects policy $p_*$  with high probability. That equilibrium involves the following behavior:  
\begin{itemize}[noitemsep]
    \item An informed loser votes $p_*$ and an informed winner votes $p^*$;
    \item An uninformed voter votes $p_*$. 
\end{itemize}
We explain why this behavior constitutes an equilibrium. As an informed voter votes for her strictly preferred policy, it suffices to consider the incentives of an uninformed voter, say Ann. Her vote affects her payoff only if it breaks a tie: of Bella and Carol, one votes $p_*$ and the other votes $p^*$. Given the stipulated behavior, Ann knows that only an informed winner would vote for $p^*$. By contrast, the vote for $p_*$ may be cast by an uninformed voter (just like Ann) or an informed loser. For $\lambda\approx 0$, Ann ascribes much higher probability to the former event. In this case, she and the other uninformed voter have an equal chance of being the remaining winner. Hence, whenever her vote is decisive, Ann believes that she is a winner with probability one-half, lower than the ex ante odds of two-thirds. At these interim odds, Ann finds it strictly optimal to vote $p_*$, ratifying that the strategy profile is a strict equilibrium.\footnote{The probability that Ann is a winner, conditional on being pivotal, is $\frac{1}{2-\lambda}$, and she has a strict incentive to vote $p_*$ if $\lambda<\frac{1}{3}$.} That is, when uninformed voters fear that support for policy $p^*$ comes from others who are better informed, they find it optimal to vote $p_*$ thereby reinforcing the initial suspicion. 

Observe that if $\lambda\approx 0$, all voters are likely uninformed. As they all vote $p_*$, that policy wins with near certainty. This outcome contrasts with what prevails in the no-information benchmark. It also contrasts with what would be chosen if all signals were realized publicly; in that setting, policy $p^*$ would win with near certainty as voters would see that they all are uninformed and therefore vote for the ex ante optimal policy. Hence, both ex ante and with public information, voters have aligned preferences and favor policy $p^*$. 

This example suggests a strategic rationale for the prevalence of zero-sum thinking even in settings where voter preferences are largely aligned. Our analysis evaluates more broadly when asymmetric information leads to such outcomes. We go beyond the example in several respects, the most important being that our general model allows uncertainty about how many voters benefit from the ex ante optimal policy. To isolate the role of asymmetric information, we start with the benchmark in which all voters share the same information and ex ante preferences; the unique equilibrium (in weakly undominated strategies) then selects the optimal policy. We compare that benchmark with an ``informationally-scarce'' setting in which each voter obtains additional private information with a small probability.\footnote{We focus on informationally-scarce settings for primarily two reasons. Conceptually, this study describes settings in which voters, being mostly uninformed, agree with high probability on which policy is better. Thus, it hones in on the  prospect for zero-sum thinking in positive-sum policy choices. We also view this focus to be reasonable from the applied perspective that it is often difficult for voters to know or predict the impact of policies. We elaborate on this motivation in \Cref{Section-Scarce,Section-CaseStudies}, and also study a setting with abundant information in \Cref{Section-PolarizationCrowding}.} 
Our analysis evaluates when the prospect of asymmetric information leads the electorate astray.

As we show, the question boils down to how an uninformed voter, say Ann, views a policy when she conditions on others learning that the policy  benefits them. Should she see the good news for others as good or bad news for herself? On one hand, others may receive good news about a policy because the policy results in many winners, which improves her odds of being a winner too. This force of advantageous selection is potent if uncertainty about the number of winners trumps distributional uncertainty, e.g., in a pure common-value election where either all voters gain or all lose from a policy. The second force is adverse selection: for any given number of winners, Ann's odds of being a winner reduce when others receive good news.
This crowding-out effect features in our example above where the number of winners is fixed.

More generally, which force dominates hinges on how a voter's payoff correlates with the signals of other voters. 
We call the collective choice problem \emph{adversely correlated} if the adverse selection effect dominates and otherwise deem it \emph{advantageously correlated}; each is a condition on primitives. Our main result (\Cref{Theorem-MainResult}) characterizes how correlation shapes equilibrium behavior in informationally-scarce settings. Stated informally, it finds: 
 \begingroup
 \addtolength\leftmargini{-0.28in}
\begin{quote}
	\textbf{Main Result.} \emph{{If the collective choice problem is adversely correlated, there is a strict equilibrium that selects the {ex ante} inferior policy with near certainty. Otherwise, every equilibrium selects the ex ante optimal policy with near certainty.}}
\end{quote}
\endgroup
Adverse correlation is thus a source of political fragility as elections may then select a policy that would not be chosen were all voters known to be uninformed. The strategic logic reflects zero-sum thinking, whereby a sufficiently large group of uninformed voters vote for the inferior policy for fear of what other voters have learned.
Not all equilibria select the inferior policy: there also exists an equilibrium in which the optimal policy wins.\footnote{For instance, in the example, there also exists an equilibrium in which all uninformed voters vote for $p^*$.} But, nevertheless, elections may not pick the right outcome and collective choice hinges on voters' ability to coordinate on a ``good'' equilibrium. By contrast, if the collective choice problem is advantageously correlated, such coordination is obviated; all equilibria---pure or mixed, symmetric or asymmetric---result in the optimal policy.\footnote{From this standpoint, our work connects to the growing interest in institutional design under adversarial equilibrium selection, as in \cite*{mathevet/perego/taneva}, \cite*{halac2021rank}, \cite*{ali2022sell}, \cite{inostroza2023adversarial}, and others.} 

For expositional clarity, we consider a model with a fixed number of voters, all of whom are ex ante symmetric, aggregating votes by simple-majority rule. But the strategic rationale for zero-sum thinking applies beyond this setting. Similar results hold for other voting rules, imposing a restriction to symmetric equilibria, and with an electorate of random size. Moreover, ex ante asymmetries only amplify the tendency for zero-sum thinking. We also show that informational scarcity is not necessary: information can be relatively abundant and yet zero-sum thinking may still push the electorate to select the inferior policy with high probability, even in a large electorate. 

The condition of adverse correlation lends itself to comparative statics. We show that policy choices that are more polarizing, in terms of the proportional loss suffered by losers relative to the gains that accrue to winners, have a greater potential to induce adverse correlation. We also formulate the ``crowding-out effect'' from learning that others are winners in terms of the standard likelihood-ratio dominance order. These results reveal features that make elections more prone to zero-sum thinking. 

Our analysis also identifies informational drivers of zero-sum thinking. If voters were to learn only about aggregate outcomes---in our model, the number of winners, but more broadly, say, aggregate GDP or economic growth---then the collective choice problem cannot be adversely correlated. By contrast, purely distributional information that identifies who is first in line to obtain gains induces adverse correlation. This contrast dovetails with analyses of how media outlets profit from selling information that polarizes rather than unifies voters \citep{martin2017bias,perego2022media}. Our results call attention to a pernicious ``downstream'' effect of this market competition on voting behavior.

We outline the remainder of our paper. \Cref{Section-Model} describes the general model and \Cref{Section-Results} our main results. \Cref{Section-UnpackingCorrelation} unpacks our correlation conditions to see what features render collective choice more adversely correlated. \Cref{Section-Extensions} considers extensions. \Cref{Section-CaseStudies} views various political episodes through the lens of our model. \Cref{Section-Conclusion} concludes. Omitted proofs are in the Appendix and Online Appendix. The rest of this introduction places our work within the context of the broader literature.

 \paragraph{Related Literature:} This paper contributes to the understanding of distrust and zero-sum thinking in politics, an issue that has been studied from various perspectives. Research in social psychology \citep*{meegan2010zero,rozycka2015belief,davidai2019politics} documents the prevalence of zero-sum thinking. Within economics, \cite{chinoy2022zero} find that both Democrat and Republican voters engage in zero-sum thinking, and that those who exhibit a greater tendency to do so also support more redistribution and stricter immigration controls.\footnote{\cite{ali2021adverse} also find evidence suggestive of zero-sum thinking in that across a range of experimental treatments on adverse and advantageous selection, subjects distrust better-informed partners who have conflicting interests but fail to trust those with aligned interests. \cite{carvalho2023zero} use an evolutionary model to show that zero-sum interactions lead to  belief systems that demotivate effort.} 
 While this body of work largely views zero-sum thinking as a psychological trait, we provide a complementary perspective that shows how this worldview emerges in settings in which rational voters worry that others have an informational advantage. 
A benefit of our approach is that it sheds light on features of a policy choice that moderate or amplify the tendency to see the world in zero-sum terms. 
%

Zero-sum thinking is closely connected to polarization and populism. Research on polarization, surveyed by \cite{iyengar2019origins}, shows that supporters of each US political party distrust members of the other party, reflecting potentially both partisan animus and the perception that policies advanced by one party are detrimental to the other. Furthermore, the work on false polarization \citep[e.g.,][]{levendusky2016mis} documents that voters overestimate their degree of polarization. Porting that finding to our context suggests that even if a collective choice problem is advantageously correlated, voters might incorrectly perceive it to be adversely correlated, which would magnify the scope for zero-sum thinking. Populism also features zero-sum thinking given that one of its defining characteristics is ``anti-elitism'' \citep{mudde2004populist,guriev2022}. Our baseline model does not distinguish between elite and non-elite voters but our extension in \Cref{Section-Elite} shows that these distinctions only amplify the strategic forces in our model. Numerous studies model populism, through the channel of signaling \citep*{acemoglu2013political}, identity politics \citep*{bonomi2021identity}, inequality aversion \citep{pastor2021inequality}, or misspecified beliefs \citep*{eliaz2020model,levy2022misspecified,szeidl2024}. Our model offers a complementary perspective, showing that asymmetric information may spur completely rational voters to distrust each other even in settings where interests are largely aligned. 

\cite{fernandez1991resistance}, an important precursor, show that a reform that benefits a majority of voters ex post may nevertheless fail ex ante if the expected payoff from the resulting lottery is not worthwhile for a majority of voters. Their model does not feature asymmetric information nor aggregate uncertainty about the number of winners; roughly put, the reasoning reflects zero-sum thinking in a zero-sum setting. In their setting, the ex ante optimal policy always wins. We show that asymmetric information amplifies the scope for political failures. The fear of what other voters have learned leads voters to choose a policy they deem ex ante inferior, i.e., zero-sum thinking in a {positive-sum} setting. 

Our work also contributes to the literature on the ``swing voter's curse,'' which studies how voting behavior is influenced by asymmetric information. \cite{feddersen/pesendorfer:96} show that, if voter interests are completely aligned, an uninformed voter might strategically abstain so as to leave the decision to better informed voters.\footnote{They also model settings in which one policy might enjoy greater support from ``partisan'' dominant-action types. They show that ``independent'' uninformed voters might counter that partisan bias by tilting their votes towards  the other policy. In this way, their analysis could rationalize uninformed voters voting for a policy they deem inferior based on all that they know. \cite{austen1996information} and \cite{feddersen1998convicting} model similar effects in pure common-value settings in which, given the prior beliefs, the voting rule tilts in favor of one policy.} In our analysis, the potential for misaligned interests implies that an uninformed voter might not defer to and, \emph{a fortiori}, might wish to act against those who are better informed.\footnote{Formally, under adverse correlation,  the equilibrium that we construct in \cref{Theorem-MainResult} would remain even if voters could abstain.}  
A few papers in this literature \citep{kim2007swing, bhattacharya2013preference,acharya2014equilibrium} show that misaligned interests could disrupt information aggregation in large elections, assuming voter types are conditionally independent.\footnote{The literature also studies other obstacles to information aggregation; see, for instance, \cite{feddersen/pesendorfer:97}, \cite{razin2003signaling}, \cite{ekmekci2020manipulated}, and \cite*{barelli2022full}.} While our work and theirs share some common threads, our focus on distributional considerations and zero-sum thinking requires a setup in which voters' type are inherently correlated. Our results also speak to the robustness and fragility of collective choice to asymmetric information rather than information aggregation per se. By highlighting how zero-sum thinking makes collective choice fragile, our analysis offers a complementary perspective on how asymmetric information shapes voting behavior. 

Our interest in fragility connects to the study of global games and higher-order beliefs  \citep{carlsson1993global,kajii1997robustness}. Closer to our work, \cite{morris2012contagious} find that small amounts of private information can disrupt asset markets through adverse selection. While related, the fragility modeled in this literature departs from that which we study. Prior work shows that higher-order uncertainty renders some equilibrium outcomes of complete information games untenable in nearby incomplete information games; i.e., the equilibrium outcome correspondence is not lower hemicontinuous. By contrast, we find that incomplete information games with scarce private information generate equilibrium outcomes that do not obtain in the complete information game. Formally, the outcome correspondence of weakly undominated equilibria is not upper hemicontinuous. As we show, second-order uncertainty alone triggers this discontinuity.

\section{Model}\label{Section-Model}
\subsection{Setup}\label{Section-Setup}
\paragraph{The Collective Choice Problem.} A group of voters $\pop\equiv\{1,\ldots,n\}$ uses simple majority rule to choose between two policies, $\pg$ and $\pb$. For simplicity, we assume that $n$ is odd so that policy $p$ wins with strictly more than   $\tau \equiv \frac{n-1}{2}$ votes.

Voters are uncertain about their payoffs from each policy. A payoff for voter $i$ is a tuple $v_i=(v^{\pg}_i,v^{\pb}_i)\in \mathcal{V}\equiv\mathcal{V}^{\pg}\times \mathcal{V}^{\pb}$, where $\mathcal{V}^p$ is a finite set of potential payoffs from policy $p$. Before casting her vote, voter $i$ receives a private signal $s_i$ drawn from a finite set of signals $\mathcal S\equiv \{s^0,\ldots,s^K\}$. A state of the world $\omega$ comprises a payoff profile $v\in \mathcal{V}^n$ and a signal profile $s\in \mathcal{S}^n$. Uncertainty is described by a probability distribution $P$ on the state space $\Omega \equiv \mathcal{V}^n\times \mathcal{S}^n$.  
The random variable $V^p$ denotes the payoff profile from choosing policy $p$ and $S$ denotes the signal profile.\footnote{In general, capital letters denote random variables and lower-case letters denote realizations. For random variable $X$, $\{x\}\equiv \{\omega : X(\omega)=x\}$ is the event where $x$ is realized, omitting the brackets when clear from the context. } The population of voters and the probability space together define the \emph{collective choice problem}.

Our baseline framework makes three assumptions. For the first assumption, we say that state $(v',s')$ \emph{permutes} state $(v,s)$ if there is a one-to-one mapping $\psi:\mathcal{N}\to\mathcal{N}$ such that $(v_i',s'_i)= (v_{\psi(i)},s_{\psi(i)})$ for all $i\in \mathcal{N}$. 

\begin{assumption}\label{Assumption-Exchangeability}
Voters are exchangeable: if $\omega'$ permutes $\omega$, then $P(\omega')=P(\omega)$.
\end{assumption}

Exchangeability focuses attention on the conflict generated by asymmetric information rather than heterogeneous ex ante preferences; we relax this assumption in \Cref{Section-Elite}. Given \Cref{Assumption-Exchangeability}, all voters agree  on the (ex ante) optimal  policy. Without loss of generality, we assume that $\pg $  is the (ex ante) optimal policy and $\pb$ is the (ex ante) inferior policy.\footnote{These policies could correspond to actual policy choices, as in a referendum, or to candidates standing in an election. For instance, policy $\pg$ may represent an incumbent politician and policy $\pb$ a challenger with little political experience.} We define the random variable $V^d\equiv V^{\pg } - V^{\pb }$ to denote the profile of payoff differences between these two policies. For a non-null event $E\subseteq \Omega$, we write voter $i$'s conditional payoff difference as
\[
V_i^d(E) \equiv \sum_{\omega \in \Omega} \Big(V_i^{\pg }(\omega)- V_i^{\pb }(\omega) \Big)P(\omega|E).
\]
If $E$ is null, we define $V_i^d(E)=0$ to simplify exposition.

Our second assumption distinguishes signal $s^0$, which is \emph{uninformative}, from the remaining signals $\inforeal\equiv \{s^1,\ldots, s^K\}$, which are \emph{informative}.

\begin{assumption}\label{Assumption-Information}
Signal $s^0$ is uninformative and every other signal is sufficient to deduce the ex post ordinal preference: for all $(v,s)\in \Omega$,
\begin{enumerate}[label=\emph{(\alph*)},ref=(\alph*),nolistsep]
\item \label{independence} if $s_i=s^0$, then $P(v,s)=P(v,s_{-i})P(s_i)$; and
\item \label{sufficiency} if $s_i\ne s^0$, then $V_i^d(s)>0$ if and only if $V_i^d(s_i)>0$.
\end{enumerate}
\end{assumption}

 \Cref{Assumption-Information}\ref{independence} asserts that signal $s^0$ realizes independently of the payoff profile and the other signals, and hence conveys no  information about these variables. \autoref{Assumption-Information}\ref{sufficiency} speaks to the informativeness of the remaining signals: for a voter who obtains an informative signal, learning the signals of others conveys no additional information about her \emph{ordinal} preference over policies. Hence, relative to the entire signal profile $s$, the private signal $s_i$ is sufficient to deduce voter $i$'s preferred policy. A special case is where the informed signal reveals directly one's cardinal payoffs from each policy, as in our introductory example. \Cref{Assumption-Information}\ref{sufficiency} goes beyond this special case by encompassing settings in which informed voters do not know all that much about their cardinal payoffs but are just well-informed \emph{relative} to the electorate about their ordinal preferences. We exploit this generality in \Cref{Section-UnpackingCorrelation}. 
  
  Among the informative signals, we say that signal $s^k$ conveys \emph{good news} if  $V_i^d(S_i=s^k)>0$ and \emph{bad news} if $V_i^d(S_i=s^k)<0$. Being ordinal, these notions convey whether an informed voter favors the ex ante optimal policy. For a signal profile $s\in \mathcal{S}^n$, $M(s)$ is the number of voters who receive informative signals, $G(s)$ is the number with good news, and $B(s)$ is the number with bad news. 
  
  Our final assumption simplifies exposition without playing a substantive role.
  
  \begin{assumption}\label{Assumption-NonRedundancy} Voters have strict preferences and both bad news and good news are possible: 
  \begin{enumerate}[label=\emph{(\alph*)},ref=(\alph*),nolistsep]
\item \label{strict} For every voter $i \in \mathcal{N}$ and non-null event $E\subseteq \Omega$, $V^d_i(E)\ne 0$. 
\item \label{non-zero} $P(B\ge 1)>0$ and $P(G\geq \tau)>0$.
\end{enumerate}
 \end{assumption}

 \autoref{Assumption-NonRedundancy}\ref{strict} obviates the need to describe how voters behave when indifferent and, with finitely many states, strict preferences are generic in the space of possible payoff profiles. \autoref{Assumption-NonRedundancy}\ref{non-zero}  rules out some trivial cases: when it is impossible to receive bad signals, voting for the inferior policy is a weakly dominated strategy; if  $P(G\geq \tau)=0$, there is always an uninteresting equilibrium where all uninformed voters select the inferior policy only because their votes are then never decisive.

  This completes the description of the baseline collective choice problem. We contrast behavior in this collective choice problem with private information to that in a \emph{public information benchmark}, where every voter observes the entire signal profile $s$.

\paragraph{Solution Concept.} A strategy for a voter is a signal-contingent probability of voting for $\pg $. We study Bayes-Nash equilibria in weakly undominated strategies, which implies that those with good news vote for $\pg $ and those with bad news vote for $\pb $. Henceforth, we refer to these as \emph{equilibria}. The following summarizes equilibrium existence in the baseline collective choice problem and the public information benchmark. 
\begin{proposition}\label{Theorem-Existence}
An equilibrium exists under both private and public information; in the latter case, the equilibrium is unique.
\end{proposition}

\subsection{Scarce Information}\label{Section-Scarce}

We primarily model settings in which information is ``scarce'': namely, all else equal, the probability that a voter obtains the uninformative signal, $s^0$, is relatively high. We describe below what scarce information entails and then briefly discuss our motivation. 

To understand how we model scarce information, suppose that signals are drawn according to the following two-stage process: first, a probability distribution chooses a state in which every voter obtains an informative signal; and second, for each voter $i$, her signal is ``deleted'' with some independent probability, which results in her obtaining the uninformative signal. For the first stage, let $P_\mathcal{I}(\cdot)\equiv P(\cdot|S\in \mathcal{M}^n)$ denote the probability distribution over states conditional on all voters receiving informative signals. For the second stage, let $\lambda\equiv P(S_i\in \mathcal{M})$ denote the probability that a voter's signal is \emph{not} deleted. \Cref{Lemma-Decomp} in the Appendix shows that the model of \Cref{Section-Model} embodies this two-stage process: under \Cref{Assumption-Exchangeability,Assumption-Information}\ref{independence}, the probability distribution $P$ decomposes into a pair $(P_\mathcal{I},\lambda)$. This decomposition permits us to vary $\lambda$ while all other variables, governed by $P_\mathcal{I}$, are held fixed. Information is then scarce if $\lambda$ is small. We define what it means for a policy to win under these circumstances.  
\begin{definition}\label{Definition-Scarce}
Policy $p$ \textit{wins in an equilibrium with scarce information} if, for every $\varepsilon$ in $ (0,1)$, there exists $\lambda_{\varepsilon}$ in $(0,1)$ such that, for all $\lambda$ in $(0,\lambda_{\varepsilon})$, policy $p$ wins with probability at least  $1-\varepsilon$ in an equilibrium of the collective choice problem $(\mathcal N, \Omega, P_{\mathcal I},\lambda)$. 
\end{definition}
\Cref{Definition-Scarce} fixes an informed collective choice problem $P_{\mathcal I}$ and studies  equilibrium outcomes of the ``diluted'' collective choice problem $(P_{\mathcal I},\lambda)$, in which a voter obtains an uninformative signal with high probability, $(1-\lambda)$, and obtains a signal according to $P_{\mathcal I}$ with a low probability, $\lambda$. 

We focus on electoral outcomes with scarce \emph{private} information, which we compare with the benchmark of scarce \emph{public} information described below. 
\begin{proposition}\label{Proposition-Public}
In the public information benchmark, policy $\pg $ wins in the unique equilibrium with scarce information. 
\end{proposition}
The logic of \Cref{Proposition-Public} is that under scarce information, all voters are likely uninformed, and in this benchmark, that event is publicly observed. Therefore, each voter casts her vote for $\pg$ thereby securing its victory. \Cref{Proposition-Public} offers the sense in which we view the policy choice to be ``positive-sum'': were all information public, all voters would likely have fully aligned interests. Against that backdrop, our main analysis evaluates how uninformed voters behave once they do not know what others have learned. 

Apart from these conceptual reasons, we also view scarce information to be relevant from an applied perspective. Many policy choices are inherently vague or complex, which makes it difficult for most voters to anticipate policy consequences. Moreover, given the recent proliferation of noisy news and media sources, each voter may be exposed to only a few information sources and not know what other voters learn.\footnote{While \Cref{Assumption-Exchangeability}\ref{independence} frames signal $s^0$ as a completely uninformative, our analysis is  compatible with settings in which each uninformed voter obtains slightly informative signals.} Our analysis speaks to how zero-sum thinking can prevail in such environments, where a majority of voters are largely uninformed but suspicious that others might have an informational advantage.

\section{The Fragility of Collective Choice}\label{Section-Results}

\subsection{Adverse Correlation}\label{Section-Correlation}

At the core of zero-sum thinking is the question: is good news for others also good news for me?
The answer depends on how a voter's payoff is correlated with others' signals.

Consider a voter, Ann, who receives the uninformative signal. Based on only her own information, the ex ante optimal remains optimal. Now suppose that she conditions on $\kappa$ voters obtaining good news and no voter obtaining bad news. Her conditional expected payoff difference is then
\begin{align*}
V^G(\kappa)\equiv V_i^d(G=\kappa,B=0,S_i=s^0),
\end{align*}
i.e., the expected payoff difference between the optimal and inferior policies conditioning on Ann receiving the uninformative signal, $\kappa$ others receiving good news, and no one receiving bad news.\footnote{We observe that $V^G$ depends only on $\mathcal{P}_{\mathcal{I}}$ and does not vary with $\lambda$.}

We elaborate on the interpretation of this term. First, this posterior expected payoff considers an event in which no voter obtains bad news; our analysis in \Cref{Section-MainResult} confirms that such events play an important role in equilibrium behavior. Second, in principle, it's unclear if Ann favors policy $\pg $ more or less once she conditions on $\kappa$ other voters obtaining good news. On one hand, the likelihood of drawing $\kappa$ good signals from $\kappa$ random draws is higher if policy $\pg $ is  likely to benefit many voters. On the other hand, for any given number of beneficiaries, Anne's chances of benefiting from $\pg $ herself is lower once $\kappa$ ``slots'' are already taken by others. These effects push in opposing directions, one bearing on aggregate considerations and the other on distributional considerations. 

The sign of $V^G(\kappa)$ varies with $\kappa$, depending on which effect dominates. 
For example, if no voter obtains good news ($\kappa=0$), the posterior expected payoff difference coincides with the prior expected payoff difference, which is strictly positive. Once some voters obtain good news, the sign depends on the relative importance of the aggregate and distributional considerations. Our introductory example features only distributional uncertainty and $V^G(\kappa)<0$ for every $\kappa>0$; knowing that anyone else obtained good news induces Ann to favor the inferior policy. By contrast, when voter preferences are perfectly aligned---as in a common-value problem---then there is only aggregate uncertainty in that all voters gain or all lose from policy $\pg $. In that case, $V^G(\kappa)>0$ for every $\kappa>0$. 

Our correlation condition assesses if $V^G$ is negative for a relevant domain.

\begin{definition}\label{Definition-Correlation}

The collective choice problem is \textit{adversely correlated} if there is some $\kappa$ in $\{1,\ldots,\tau\}$ such that $V^G(\kappa)<0$. Otherwise, the collective choice problem is \textit{advantageously correlated}.\footnote{By \Cref{Assumption-NonRedundancy}\ref{strict}, advantageous correlation equates to $V^G(\kappa)>0$ for every $\kappa$ in $\{1,\ldots,\tau\}$.}

\end{definition}
Recall that $\tau\equiv (n-1)/2$ is the number of votes for policy $\pg $ that makes a voter pivotal. The criterion above evaluates if there is any coalition of weakly smaller size whose good news, on net, makes an uninformed voter favor the inferior policy $\pb $. 
This criterion jointly evaluates the payoff and information structure: as we show in \Cref{Section-UnpackingCorrelation}, fixing a distribution of payoffs, some information structures render the collective choice problem adversely correlated whereas others make it advantageously correlated. Likewise, fixing an information structure,  payoffs can determine whether the collective choice problem is adversely or advantageously correlated.

\subsection{Main Result}\label{Section-MainResult}

We show that the fragility of collective choice hinges on its correlation. 
\begin{theorem}\label{Theorem-MainResult}
The inferior policy $\pb $ wins in a strict equilibrium with scarce information if the collective choice problem is adversely correlated. Otherwise the optimal policy $\pg $ wins in every equilibrium with scarce information. 
\end{theorem}

Recall that were each voter known to be uninformed, or if all information were public and scarce, every equilibrium would select policy $\pg$: without private information, the electorate faces a positive-sum policy choice in which the optimal policy receives unanimous support. \Cref{Theorem-MainResult} shows, by contrast, that the slightest prospect of private information could result in a different equilibrium outcome. 
Voting behavior is fragile to private information whenever the collective choice problem is adversely correlated: formally, the outcome correspondence of weakly undominated equilibria violates upper hemicontinuity as $\lambda\rightarrow 0$. Conversely, advantageous correlation assures robustness. Not only does the optimal policy $\pg $ win with non-trivial odds, it does so with near certainty in any equilibrium, pure or mixed, asymmetric or symmetric.\footnote{An equivalent way to frame our results is through the lens of public and private information. Suppose that all voters obtain noisy public information that indicates that policy $\pg $ is superior. If it were commonly known that no voter obtains any additional private information, the electorate would use this public information to select policy $\pg $ in every weakly undominated equilibrium. However, if some voters obtain additional private information with even the slightest chance, under adverse correlation, there is an equilibrium in which the electorate fails to use public information to select the better policy. Thus, a better informed electorate might make worse choices when some of that better information is received privately by voters.}

To see how adverse correlation results in policy $\pb $ winning the election, we start with the simplest case, namely $V^G(\tau)<0$. Then, a strict equilibrium akin to that of our introductory example implements the inferior policy: all uninformed voters vote for policy $\pb $. An uninformed Ann infers, conditional on being pivotal, that $\tau$ voters have voted for policy $\pg $. By construction, each of these voters must have heard good news. As information is scarce, Ann believes all those who voted for $\pb $ are likely uninformed, i.e., the number of voters who received bad news is likely $0$. Her expected payoff difference conditional on being pivotal then approximates $V^G(\tau)<0$, which implies that she strictly prefers to vote for policy $\pb $. Because most voters are uninformed, it wins with high probability.

Other cases require more elaborate constructions. Let $V^G(\tau)>0$ and $\kappa^*$ be the highest value of $\kappa$ in $\{1,\ldots,\tau\}$ such that $V^G(\kappa)<0$. We construct a (strict) pure-strategy equilibrium that is asymmetric. In this equilibrium, we label some voters as ``suspicious'' and all others as ``sanguine.'' A suspicious (resp., sanguine) voter casts her vote for policy $\pb $ (resp., $\pg $) when she is uninformed. We specify that $(\tau-\kappa^*)$ voters are sanguine and all others are suspicious. Our argument establishing that this is an equilibrium hews to the following logic:
\begin{itemize}
	\item A suspicious uninformed voter, conditional on being pivotal, places high odds on the event where (i) of the $\tau$ votes for policy $\pg $, $(\tau-\kappa^*)$ are cast by uninformed sanguine voters and the remaining $\kappa^*$ votes are cast by suspicious voters who obtained good news; (ii) all the $(n-\tau-1)$ votes for policy $\pb $ are cast by uninformed suspicious voters. Hence, her conditional expected payoff difference is close to $V^G(\kappa^*)<0$. Thus, she strictly prefers to vote for policy $\pb $. 
	\item A sanguine uninformed voter makes a different calculation that stems from her being sanguine: conditional on being pivotal, she puts high odds that of the $\tau$ votes for policy $\pg $, $(\tau-\kappa^*-1)$ are cast by uninformed sanguine voters and $(\kappa^*+1)$ votes are cast by suspicious voters who obtained good news. Hence, her conditional expected payoff difference is close to $V^G(\kappa^*+1)$, which is strictly  positive as $\kappa^*$ is the highest value of $\kappa$ such that $V^G(\kappa)<0$. Thus, she strictly prefers to vote for policy $\pg $. 
\end{itemize}

Having argued that the strategy profile is an equilibrium, we note that as $\lambda\rightarrow 0$, the election is likely decided by the uninformed voters. Given that sanguine voters are in a minority, the inferior policy $\pb $ wins with near certainty. 

\Cref{Theorem-MainResult} also asserts that under advantageous correlation, the optimal policy prevails in \emph{every} equilibrium with near certainty. Hence adverse correlation is also necessary for electoral failures. This direction is considerably more challenging to prove given the large number of candidate equilibria, including those that are asymmetric and in mixed strategies. The key idea, we show, is that as private information becomes increasingly scarce ($\lambda\to 0$), the relevant posterior payoff of an uninformed voter approximates a weighted sum of the $V^G(\kappa)$ for different values of $\kappa\in \{0,..,\tau\}$ for  \emph{any} weakly undominated strategy profile. The specific weights depend on the strategy profile but, regardless of the weights,  advantageous correlation implies that the sum is strictly positive  because $V^G(\kappa)>0$ for \emph{every} $\kappa$. Hence, the decisions of others in any equilibrium never generates enough bad news to sway uninformed voters to vote for policy $\pb $ with non-trivial probability.

In light of our main result, the reader may wonder if adverse correlation nevertheless accommodates a ``good'' equilibrium that selects the optimal policy? Yes.
\begin{theorem}\label{Theorem-GoodEquilibrium}
Regardless of correlation, policy $\pg $ wins in at least some equilibrium with scarce information.
\end{theorem}
The logic of \Cref{Theorem-GoodEquilibrium} is that although the information structure is a primitive of the collective choice problem, how an uninformed voter interprets others' votes is determined in equilibrium. Our construction in \Cref{Theorem-MainResult} sways an uninformed Ann towards policy $\pb $ by having most votes in favor of policy $\pg $ to be cast by those who have good news so that, under adverse correlation, policy $\pg $ is selected adversely to her interests. By the same logic, a strategy profile in which most votes in favor of policy $\pb $ are cast by those who have bad news could make policy $\pb $ selected adversely to Ann's interests, which would push Ann to vote for policy $\pg $. Our proof of \Cref{Theorem-GoodEquilibrium} shows that the ex ante inferiority of policy $\pb $ implies that it is always feasible to construct such an equilibrium. As uninformed voters overwhelmingly vote for policy $\pg $ in that equilibrium, it wins when information is scarce.

\Cref{Theorem-GoodEquilibrium} suggests that fragility is not a concern \emph{if} voters can coordinate on equilibria that maximize ex ante welfare. We take the perspective that presuming voter coordination is not modest, particularly in large electorates or in the absence of an ex ante stage at which all voters are known to be symmetrically informed. \Cref{Theorem-MainResult} implies that, under adverse correlation, voter coordination is \emph{necessary} to avoid an inferior outcome that is rationalized by neither voters' information nor their preferences; the prospect of asymmetric information, however slight, can result in this perverse outcome. By contrast, advantageous correlation obviates voter coordination as the optimal policy is assured across all equilibria.

Our analysis applies beyond our baseline setting. We assume simple-majority rule but our analysis invokes it only for simplicity. Suppose that policy $\pg $ passes if it obtains at least $\tau+1$ votes where $\tau$ is now any number in $\{1,\ldots,n-1\}$. We would continue to define adverse and advantageous correlation as per \Cref{Definition-Correlation}, and \Cref{Theorem-MainResult} still holds. Therefore, increasing the number of votes required for policy $\pg $ to pass only expands the set of parameters for which adverse correlation holds. 
 
 \Cref{Section-Extensions} describes elaborations that require further analysis. A similar characterization obtains for symmetric equilibria under a slightly different correlation condition; that analysis also accommodates population uncertainty. We also relax the condition that voters are ex ante symmetric, allowing for some voters to be ``elites'' who are more likely to benefit from the optimal policy. We show that this ex ante conflict only amplifies the forces described here. Before describing these extensions, we focus on a tractable class of the baseline model to identify various sources of adverse correlation.

\section{Sources of Adverse Correlation}\label{Section-UnpackingCorrelation}

In this section, we study features of a collective choice problem that lead to adverse correlation. 
For this comparison, we focus on a class of  collective choice problem with \emph{binary payoffs}: $V^d_i \in \{\vw,-\vl\}$ for some $\vw$ and $\vl$ strictly positive, so that each voter's ex post payoff from $\pg$ is either $\vw$ higher or $\vl$ lower than that from $\pb$.\footnote{For instance, $\Omega = \{\vw,-\vl\}^n\times \{0\}^n \times \mathcal{S}^n$ for some $\vw>0$ and $\vl>0$.} The event $W_i \equiv \{\omega : V_i^d(\omega)>0\}$ are all states where voter $i$ gains from policy $\pg $, and is referred to as a \emph{winner}. The number of winners in state $\omega$ is $W(\omega)\equiv |\{i\in  \mathcal{N}:\omega\in W_i \}|$. As policy $\pg $ is ex ante optimal, we are assuming throughout that $P(W_i)\vw -(1-P(W_i))\vl >0$.

In this class of collective choice problems, we study how varying the payoffs, the probability distribution on the number of winners, and the nature of the information conveyed by signals affects the correlation structure. 

\subsection{Polarization Ratios and the Crowding-Out Effect}\label{Section-PolarizationCrowding}

We first fix a simple information structure to focus on payoffs: suppose that informative signals convey \emph{perfect news} in that an informed voter learns if she is a winner or loser, i.e., $P(W_i|S_i =s^k)\in \{0,1\}$ for each informative signal $s^k$. The collective choice problem can then be represented through the tuple $(P_W,v)$ where $P_W$ is the (marginal) distribution on the number of winners, $P_W:\{0,\ldots,n\}\rightarrow [0,1]$, and $v\equiv(\vw,\vl)$ specifies the ex post payoff differences. We write $V^G(\kappa;P_W,v)$ as the value of $V^G(\kappa)$ for this collective choice problem, and order collective choice problems by their propensity for adverse correlation. 
\begin{definition}\label{Definition-AdverselyCorrelated-2}
	A collective choice problem $(P_W,v)$ is \textit{more adversely correlated} than the collective choice problem $(P_W',v')$ if, for every $\kappa \in \{1,\ldots,\tau\}$, 
	\begin{align*}
		V^G(\kappa|P'_W,v') <0 \implies V^G(\kappa|P_W,v)<0.
	\end{align*}	
We denote this binary relation by $\succeq_{AC}$.\footnote{We note that $\succeq_{AC}$ is a preorder in that it is reflexive and transitive but not complete.}
\end{definition}

We identify two sources of adverse correlation. The first views payoffs through their \emph{polarization ratio}: given binary payoffs $(\vw,\vl)$, the \emph{polarization ratio} $\vl/\vw$ measures the proportional costs incurred by losers relative to the gains that accrue to winners. The second views the crowding-out effect through likelihood ratios: the probability distribution $P_W'$ likelihood-ratio dominates $P_W$, denoted $P_W'\succeq_{LR} P_W$, if $P'_W(w')P_W(w) \geq P'_W(w)P_W(w')$ whenever $w'>w$.
\begin{proposition}\label{Theorem-ComparativeStatics}
The following comparative statics results hold:
\begin{enumerate}[label=\emph{(\alph*)},ref=(\alph*),nolistsep]
	\item A higher polarization ratio induces more adverse correlation: $ (P_W,v)\succeq_{AC} (P_W,v')$ if $\vl/\vw \geq \vl'/\vw'$.
	\item Likelihood-ratio dominance induces less adverse correlation: $(P_W',v)\succeq_{AC} (P_W,v)$ if $P_W\succeq_{LR} P_W'$. 
\end{enumerate} 
\end{proposition}
A higher polarization ratio increases adverse correlation because, relative to the inferior policy $\pb $, a higher polarization ratio worsens the gamble from the optimal policy $\pg $ both ex ante and ex interim. By contrast, a likelihood-ratio dominant increase in the marginal distribution over the number of winners has the opposite effect because it reduces crowding-out from learning that others received good news. In combination with \Cref{Theorem-MainResult}, \Cref{Theorem-ComparativeStatics} implies that increasing the polarization ratio or reducing the likelihood ratio makes collective choice more fragile to asymmetric information.

An implication of \Cref{Theorem-ComparativeStatics} is that redistributive programs may alleviate zero-sum thinking. Coupling policy $p^*$ with transfers from winners to losers would lower the polarization ratio and mitigate the concern of crowding-out. Concretely, if trade expansion is accompanied by retraining programs for displaced workers, then a referendum on a trade agreement may be more likely to pass as the collective choice problem would then be less adversely correlated. Redistribution can transform an adversely correlated collective choice problem into one that is advantageously correlated.

\paragraph{A Large Election Analysis:} We build on this analysis of crowding-out behavior to show that zero-sum thinking may prevail even when information is abundant. For this analysis, we hold the primitives fixed and study behavior as the population size $n\rightarrow \infty$, assuming throughout that $n$ is odd. Our focus here is on distributional uncertainty and the crowding-out effect so, as in the introductory example, we fix the proportion of winners: there is some $q$ in $\left[\sfrac{1}{2},1\right)$ such that in the election with $n$ voters, there are $\ceil{qn}$ winners. Policy $\pg $ being ex-ante optimal for every $n$ is then equivalent to the polarization ratio, $v_\ell/v_w$, being strictly less than $q/(1-q)$.

We identify a critical polarization ratio that determines if adverse or advantageous correlation holds in a large election: using $\{\tau_n\}$ to denote the event where $\frac{n-1}{2}$ voters receive good news and no voter receives bad news, the critical polarization ratio is
\begin{align*}
\rho\equiv  \lim_{n\rightarrow \infty}\frac{P_n(W_i|\{\tau_n\})}{1-P_n(W_i|\{\tau_n\})}= \frac{q-\frac{1}{2}}{1-q}.
\end{align*}
Our first result connects this critical polarization ratio to adverse correlation in large elections.
\begin{proposition}\label{Proposition-LocalAdverseCorrelation}
    If $\vl/\vw> \rho$, then there exists $N$ such that for all $n>N$, the $n$-voter collective choice problem has adverse correlation; otherwise, for every $n$, the $n$-voter collective choice problem has advantageous correlation.
\end{proposition}
\Cref{Proposition-LocalAdverseCorrelation} obtains a necessary and sufficient condition for adverse correlation in all sufficiently large elections. We use this result to characterize electoral outcomes in a large election.
\begin{proposition}\label{Proposition-LargePopulation}
If $\vl/\vw> \rho$, there exists $\lambda^*>0$ such that, for all $\lambda <\lambda^*$, the inferior policy wins in a strict equilibrium of a large election.\footnote{Formally, for every $\epsilon$ in $(0,1)$, there exists $N_{\epsilon}$ such that, for all $n>N_{\epsilon}$, there is a strict equilibrium of the $n$-voter collective choice problem in which the inferior policy wins with probability at least $1-\epsilon$.}
    Otherwise, for every population size $n$ and $\lambda$ in $(0,1)$, the optimal policy wins with probability $1$ in every equilibrium.
\end{proposition}

Hence, if there is adverse correlation, the inferior policy wins with near certainty in a strict equilibrium of a large election; otherwise, the optimal policy wins with certainty in every equilibrium for every population size. 

We note several features of this result. First, it fixes a $\lambda>0$ and then takes $n\rightarrow \infty$. The order of limits therefore differs from that of  \Cref{Theorem-MainResult}. Information need not be scarce and may actually be abundant. We show in the proof that $\lambda^* \ge 1- \left(v_\ell/v_w\right)^{-1}\rho$. Using the parameters of the introductory example, namely $q=\sfrac{2}{3}$, $v_\ell = 3$, and $v_w=2$, the inferior policy wins in a strict equilibrium of a large election for any $\lambda \in (0,\sfrac{2}{3})$. Hence, it may even be more likely that voters are informed than uninformed, and nonetheless, the inferior policy wins. Second, as the bound $1- \left(v_\ell/v_w\right)^{-1}\rho$ indicates, increasing the polarization ratio only expands the region in which the inferior policy  wins. Third, fixing all other parameters, increasing the population size may expand the range of $\lambda$ for which there is an equilibrium in which all uninformed voters vote for $p_*$. This effect is evident in our introductory example: for $n=3$, this equilibrium exists if $\lambda$ is in $(0,\sfrac{1}{3})$ whereas if $n$ is sufficiently large, the equilibrium exists whenever $\lambda$ is in $(0,\sfrac{2}{3})$. Finally, a large election makes it even more likely that policy $p_*$ wins in a large election.\footnote{To illustrate, consider the introductory example and suppose that $\lambda=\sfrac{1}{4}$. Then, if $n=3$, the probability that $p_*$ wins in our constructed equilibrium is $\sfrac{27}{32}$ whereas this probability converges to $1$ as $n\rightarrow \infty$.} Thus, in a large election with abundant information, zero-sum thinking may not only prevail but also be even more pronounced.

\subsection{Aggregate and Distributional Information}\label{Section-NatureInformation} 
Electoral choices are also shaped by the kind of information voters obtain. Contrast the following messages by Republican Presidents about the economic consequences of trade liberalization:

\begin{quote}
    \textit{``Free trade serves the cause of economic progress\ldots''} -- Reagan, 1982\medskip

    \textit{``Members of the club---the consultants, the pollsters, the politicians, the pundits, and the special interests---grow rich and powerful while the American people grow poorer and more isolated\ldots"}\hspace{.53in} -- Trump, 2016
\end{quote}

The former emphasizes aggregate consequences without referring to who benefits and loses whereas the latter focuses on those distributional consequences without speaking about aggregate gains. Politicians and media outlets can choose between focusing their messaging on aggregate or distributional consequences. Here, we identify how this choice affects voting behavior. 

To this end, we depart from the perfect news considered in \Cref{Section-PolarizationCrowding}. Instead, we say that signals \emph{convey only aggregate news} if $P(V=v|S=s,W=w)=P(V=v|W=w)$ for any payoff profile $v$, signal profile $s$, and number of winners $w$ such that $P(W=w)>0$. That is, conditioning on the number of winners, voters learn nothing further about the payoffs from their signals.
\begin{proposition}\label{Proposition-Aggregate}
Suppose signals convey only aggregate news. Then for every polarization ratio, the collective choice problem is advantageously correlated.
\end{proposition} 
 
We contrast this case with one where signals \emph{convey only distributional news}: $P(W=w|S=s)=P(W=w)$ for any signal profile $s$ and number of winners $w$, i.e., signals are uninformative about the number of winners.
\begin{proposition}\label{Proposition-Distributional}
	Suppose signals convey only distributional news. Then holding the signal structure fixed, there is a polarization ratio $\hat{\rho}$ such that the collective choice problem is adversely correlated whenever $\vl/\vw \geq \hat{\rho}$.\footnote{In the proof, we show that this critical $\hat{\rho}$ is low enough that policy $\pg $ remains optimal.} 
\end{proposition}
 The contrast between \Cref{Proposition-Aggregate,Proposition-Distributional} reveals how the information structure may preclude or induce adverse correlation by focusing on aggregate or distributional news. This finding dovetails with \cite{martin2017bias} and \cite{perego2022media} who show that media providers may profit from delivering information that polarizes voters. In conjunction with their   analysis, \Cref{Proposition-Distributional} suggests that this profit-seeking behavior may have detrimental effects on elections.\footnote{Focusing on a different force, \cite{yuksel2022specialized} shows that segregation in news consumption has a further polarizing effect on the electoral platforms chosen by parties.}

\section{Extensions}\label{Section-Extensions}
\subsection{A Characterization for Symmetric Equilibria}\label{Section-Symmetry}
 
 The analysis of \Cref{Theorem-MainResult} considers all equilibria, including those that are asymmetric.\footnote{We note that the equilibria described in \Cref{Theorem-Existence}-\ref{Proposition-Public} and \Cref{Theorem-GoodEquilibrium} are symmetric.} Restricting to symmetric equilibria, a stronger form of adverse correlation is necessary for the inferior policy to win with scarce  information, while a weaker form of advantageous correlation is sufficient to guarantee that the optimal policy wins across all symmetric equilibria.
 
In a symmetric (weakly undominated) equilibrium, informed voters choose their preferred policy and all uniformed voters choose $\pg $ with the same probability $\alpha \in [0,1]$.  When $\alpha \in (0,1)$, a voter can be pivotal when $\kappa$ others receive good news for any $\kappa\in \{0,...,\tau\}$, because it is always possible for exactly $\tau-\kappa$ uninformed voters to choose $\pg $ along with the $\kappa$ voters who received good news. However, these events are not equally likely, depending both on the primitive probability of events and the behavior of uninformed voters. The relevant correlation measure for symmetric equilibria therefore involves a weighted sum of the conditional payoffs $V^G(\cdot)$ evaluated at different values of $\kappa\in \{0,...,\tau\}$:
 \begin{align*}
  \Kappa(\theta) =\sum_{\kappa=0}^{\tau}\theta^{\kappa}\binom{\tau}{\kappa}P(G=\kappa|B=0)  V^G(\kappa).
 \end{align*}
This sum again focuses on the perspective of an uninformed voter, whose expected payoff difference conditional on $\kappa$ others receiving good news and no one receiving bad news is $V^G(\kappa)$. In each summand, $P(G=\kappa|B=0)$ is the primitive probability that $\kappa$ voters receive good news, conditional on no one receiving bad news. The binomial coefficient is an adjustment factor to account for the number of ways that the uninformed voter can be pivotal when no one receives bad news.\footnote{Note that $\binom{\tau}{\kappa}=\binom{n}{\tau}^{-1}\binom{n}{\kappa}\binom{n-\kappa}{\tau-\kappa}$, where $\binom{n}{\kappa}$ is the number of ways of selecting the $\kappa$ voters that receive good news, $\binom{n-\kappa}{\tau-\kappa}$ is the number of ways of selecting $\tau-\kappa$ uninformed voters to vote for the optimal policy along with voters who received good news, and their product is normalized by the total number of ways of selecting $\tau$ votes for the optimal policy $\binom{n}{\tau}$. } As we elaborate below, the variable $\theta$ encodes the relative likelihood that, in a symmetric mixed strategy profile, a vote for policy $\pg $ is cast by an informed rather than an uninformed voter, when $\kappa$ voters received good news and no one received bad news. In particular, $\theta^{\kappa}$ offers additional flexibility in weighting the summands: taking $\theta\rightarrow 0$ concentrates weight on the term that involves $V^G(0)$ whereas $\theta\rightarrow\infty$ focuses the sum on the term that involves $V^G(\tau)$. 

The relevant \emph{correlation measure} here is the infimum, $\Kappa_*\equiv \inf_{\theta\in \Re_{++}} \Kappa(\theta)$.
\begin{definition}\label{Definition-SymmetricCorrelation}
The collective choice problem is \textit{strongly adversely correlated} if $\mathcal{K}_*<0$ and \textit{weakly advantageously correlated} if $\mathcal{K}_*>0$. 
\end{definition}

On one hand, $V^G(\kappa)>0$ for all $\kappa\in \{0,...,\tau\}$ is sufficient but not necessary for $\Kappa_*>0$, and so advantageous correlation implies weak advantageous correlation but not vice versa. On the other hand, $V^G(\kappa)<0$ for some $\kappa\in \{0,...,\tau\}$ is necessary but not sufficient for $\Kappa_*<0$, and so strong adverse correlation implies adverse correlation but not vice versa.\footnote{However, $V^G(\tau)<0$ does imply $\Kappa_*<0$ because, as $\theta\to \infty$, the $\tau$-th summand in $\Kappa(\theta)$ dominates.}

Every collective choice problem is either strongly adversely or weakly advantageously correlated except in the knife-edge case where $\Kappa_*=0$, which is possible only on a measure-$0$ set of parameters. Parallel to \Cref{Theorem-MainResult}, the following theorem therefore essentially characterizes when a collective choice problem is fragile to asymmetric information with a restriction to symmetric equilibria.
 \begin{theorem}\label{Theorem-Symmetric}
 If the collective choice problem is strongly adversely correlated, the inferior policy $\pb $ wins in a symmetric equilibrium with scarce information. By contrast, if the collective choice problem is weakly advantageously correlated, then the optimal policy $\pg $ wins in every symmetric equilibrium with scarce information. 
 \end{theorem}
Hence, fragility also emerges with symmetric equilibria. However, establishing this fragility requires a different argument. Suppose that the collective choice problem is strongly adversely correlated; moreover, assume that $V^G(\tau)>0$.\footnote{If $V^G(\tau)<0$, the equilibrium constructed in the proof of \Cref{Theorem-MainResult} is already symmetric.} We show that a sequence of symmetric mixed strategy equilibria selects policy $\pb $ with probability approaching $1$ as $\lambda \to 0$. The idea is that $\Kappa_*<0$ implies the existence of $\tilde\theta$ such that $\Kappa(\tilde\theta)=0$. For each $\lambda\in (0,1)$, we then choose $\alpha$ such that $\tilde\theta = \frac{\lambda}{(1-\lambda)\alpha }$; in other words, $\tilde \theta$ is the relative likelihood that a vote for $\pg $ is cast by an informed voter. Choosing $\alpha$ at this rate guarantees that an uninformed voter is indifferent between policies $\pg $ and $\pb $ conditional on being pivotal, which rationalizes her mixing. Moreover, as $\lambda\rightarrow 0$, the probability that an uninformed voter votes for $\pg $ converges to $0$. As most voters are uninformed, policy $\pb $ then wins with near certainty.

The converse also requires a different argument because weak advantageous correlation does not imply advantageous correlation. However, weak advantageous correlation is sufficient to ensure that, regardless of how one pools votes for $\pg $ from uninformed voters and those who hear good news, uninformed voters are never swayed to vote for policy $\pb $ with substantial probability in any symmetric equilibrium. 

 \subsection{Population Uncertainty}\label{Section-Population}

The characterization for symmetric equilibria can also accommodate population uncertainty.
Suppose that, as in \cite{myerson1998population}, the population size is random. For simplicity, we assume that the population size is always odd and a policy $p\in \{p^*,p_*\}$ is then chosen by simple majority rule: for a realized population of $n$ voters, policy $p$ wins if it receives at least $\tau(n)+1$ votes, where $\tau(n)\equiv \frac{n-1}{2}$. We make the following assumption about the random population size. 

\begin{assumption}\label{Assumption-RP}
The population size $N$ is drawn from a probability measure $Q$ with a finite expectation and support $\mathcal{Q}$, which is a subset of the odd positive integers strictly greater than one.
\end{assumption}

Let $\Omega^n = \mathcal{V}^n\times \mathcal{S}^n$ be the state space for a realized population size $n$. Uncertainty is described by the random population size  $(\mathcal{Q},Q)$ and a stochastic process $\{ (\Omega^n,P_n):n\in \mathcal{Q}\}$, where $P_n$ is a probability distribution over payoff and signals profiles, $\omega\in \Omega^n$, for the voters in an election for population size $n$. We adapt our main assumptions in \Cref{Section-Model} to apply conditional on each population size $n$ (see online appendix for a formal description). In addition, we assume that for every population size $n$, the ex-ante optimal policy $\pg $ is also optimal conditioning on only the population size $n$.

We generalize the correlation measure considered in \Cref{Section-Symmetry}, considering inferences that a voter draws were she to think that the population size is $n_0$, which is the smallest population size in $\mathcal{Q}$. Adapting our previous notation, let $V^G(\kappa,n_0)\equiv V^d_i(S_i=s^0,G=\kappa,B=0,N=n_0)$
denote the expected payoff difference for a voter who receives the uninformative signal, conditioning on $\kappa$ voters obtaining good news, no voter obtaining bad news, and there being $n_0$ voters (where $n_0\geq \kappa$). This term feeds into the relevant correlation measure:
\begin{align*}
    \mathcal{K}_*(n_0) \equiv \inf_{\theta\in \Re_{++}}\sum_{\kappa=0}^{\tau(n_0)} \theta^{\kappa} \binom{\tau(n_0)}{\kappa} P(G=\kappa|B=0)V^G(\kappa,n_0).
\end{align*}

\begin{theorem}\label{Theorem-PopulationUncertainty}
    If $\mathcal{K}_*(n_0)<0$, the inferior policy $\pb $ wins in a symmetric equilibrium with scarce information. By contrast, if $\mathcal{K}_*(n_0)>0$, the optimal policy $\pg $ wins in every symmetric equilibrium with scarce information.
\end{theorem}
The key idea is that an uninformed voter believes she is most likely to be pivotal in a small election, and increasingly so as information becomes scarce. Hence, although she may expect that the population is large, her beliefs about adverse correlation at population size $n_0$ drive behavior.

 \subsection{The Role of Elites}\label{Section-Elite}

 Our baseline model isolates the role of asymmetric information by assuming away all other differences between voters. However, many policy debates that feature zero-sum thinking also involve voters who are ex ante heterogeneous. In this section, we show that heterogeneity makes the election more ripe for zero-sum thinking. 

We continue to assume that for each voter, $\pg$ is ex ante optimal. We weaken \Cref{Assumption-Exchangeability} to a ``within-group exchangeability'' notion: we decompose voters into ``elites'' and ``non-elites'' and assume that voters are exchangeable within each group but not necessarily across groups. The collective choice problem therefore consists of the set of elite voters $\mathcal{E}$, non-elite voters $\mathcal{NE}$, and a probability space $(\Omega,P)=(\mathcal{V}^n\times \mathcal{S}^n,P)$ such that \Cref{Assumption-Information,Assumption-NonRedundancy} are satisfied, and for any permutations, $\psi_E:\mathcal{E}\to \mathcal{E}$ and $\psi_N:\mathcal{NE}\to\mathcal{NE}$, and state $\omega$,$P(\omega_{\mathcal{E}},\omega_{\mathcal{NE}})=P(\omega_{\psi_E(\mathcal{E})},\omega_{\psi_N(\mathcal{NE})})$.

Given a non-empty set of voters $\mathcal{H}$, let $G_{\mathcal{H}}$ denote the random variable describing the number of voters in $\mathcal{H}$ who received good news.
\begin{definition}\label{Definition_MajNegcor}
The collective choice problem is \textit{elite-adversely correlated} if there exists a binary partition of the electorate, $\{\mathcal{E},\mathcal{NE}\}$ such that the following hold:
\begin{enumerate}[label=\emph{(\alph*)},ref=(\alph*),nolistsep]
	\item \label{Elites-minority} Elites are a minority: $|\mathcal{E}|< \trule$. 
	\item \label{Elites-nofear} Elites do not fear the support of others: 
	\begin{align*}
	V_i^d(S_i=s^0,B=0,G=G_{\mathcal{NE}}=\tau-|\mathcal{E}|+1)>0\text{ for every }i\in \mathcal{E}.	
	\end{align*}
	\item \label{Nonelites-fear} Non-elites fear the support of others: 
	\begin{align*}
	V_i^d(S_i=s^0,B=0,G=G_{\mathcal{NE}}=\tau-|\mathcal{E}|)<0\text{ for every }i\in \mathcal{NE}.	
	\end{align*}
\end{enumerate}
\end{definition}
\Cref{Definition_MajNegcor}\ref{Elites-minority} asserts that the elites are a minority of the electorate. Part \ref{Elites-nofear} states that these voters continue to support the optimal policy even after conditioning on the support of others. In other words, they are not concerned about adverse selection. Part \ref{Nonelites-fear} states that non-elites, by contrast, {are} concerned by adverse selection: knowing that all elites vote for the optimal policy, each views her odds of gaining from $\pg $ to be low when sufficiently many non-elite voters obtain good news. Parts \ref{Elites-nofear} and \ref{Nonelites-fear} together imply that elites are more likely to benefit from policy $\pg $. 

A special case of \Cref{Definition_MajNegcor} is where each elite voter has a higher ``rank'' than every non-elite voter in that an elite voter is guaranteed to gain from the optimal policy $\pg $ whenever a non-elite voter does. An elite voter then is elated to learn that any non-elite voter has good news for it assures that she too gains from $\pg $. By contrast, non-elites are crowded out from being winners both by elites and other non-elites. In this vein, \Cref{Definition_MajNegcor} views the optimal policy as a gamble that is simply more likely to benefit elites before its rewards trickle down to non-elite voters. 

As voters agree that $\pg $ is ex-ante optimal, it remains the winner with scarce \emph{public} information. But collective choice may still be fragile to private information.
\begin{proposition}\label{Prop-MajNegcor}
If payoffs are elite-adversely correlated, the inferior policy $\pb $ wins in a strict equilibrium with scarce information. 
\end{proposition}

The idea is that elite voters, unconcerned by adverse selection, vote for $\pg $ even when they are uninformed. As elites are a minority, $\pg $ can only be in the race if it has sufficient support from non-elites. An uninformed non-elite voter then worries about being crowded out and hence has a strict incentive to vote for $\pg $.  

Moreover, the presence of elite voters can exacerbate adverse selection for non-elites. To see how, we specialize to the binary payoff setup described in \Cref{Section-PolarizationCrowding}, in which informed voters obtain perfect news, and assume that $W_i \subseteq W_j$ for all $i\in \mathcal{NE}$ and $j\in \mathcal{E}$, so that an elite voter is assured to gain from $\pg$ if a non-elite voter does. The collective choice problem is then fully described by the probability that a voter receives an informative signal $\lambda$, the ex-ante distribution over the number of winners $P_W$, the payoffs $v=(v_W,v_L)$, and the number of elite voters $|\mathcal{E}|\equiv e$. 
Let $\tilde{E}(e)=\{\omega\in \Omega: S_i=s^0, B=0, G=G_{\mathcal{NE}}=\tau-e\}$. 

\begin{proposition}\label{Prop-EliteComparative}
For a non-elite voter $i$, the conditional expected payoff $V_i^d(\tilde{E}(e) |P_W,v,e)$ is strictly decreasing in the number of elites $e$.
\end{proposition}
Hence, increasing the size of the elite group expands the range of polarization ratios for which the collective choice problem is majority-adversely correlated, leading to a greater scope for political failures.

\section{Zero-Sum Thinking in Practice}\label{Section-CaseStudies}

Here, we view some recent political episodes through the lens of our model.

Consider, first, the public debate on whether the US government should bail out financial institutions following the 2008 subprime mortgage crisis. The Bush administration and other government officials emphasized that a bailout was necessary:  Ben Bernanke, the then Chairman of the Federal Reserve, argued in a Senate hearing on 24th September 2008 that
\begin{quote}
   \textit{``I believe if the credit markets are not functioning, that jobs will be lost, the unemployment rate will rise, more houses will be foreclosed upon, GDP will contract, that the economy will just not be able to recover in a normal, healthy way, no matter what other policies are taken.''}
\end{quote}
Nevertheless, the bill proposing the Emergency Economic Stabilization Act (EESA) did not pass in the US House of Representatives in an initial vote on 29th September. Congressional leaders were influenced by the large-scale protests against the perceived zero-sum transfer from Main Street to Wall Street, a popular sentiment that was echoed by Democratic presidential candidate Barack Obama's refrain that “taxpayers shouldn’t be spending a dime to reward CEOs on Wall Street while they’re going out the door.” The EESA passed a second vote on October 3rd only after one of the biggest one-day market collapses in US history sent a visible signal of the severe macroeconomic risks from a failure to stabilize the financial system.

While it was clear that the financial industry would profit from the proposed $\$700$ billion bailout, there was considerable aggregate and distributional uncertainty about the broader general equilibrium effects of the deteriorating credit market conditions. Would a bailout be solely distributional, transferring wealth from taxpayers to CEOs, or would there be significant beneficial aggregate effects of bailing out financial institutions? Information on this score was scarce given the sheer complexity of predicting the effects of approving or rejecting the bill, and much of the information in the public sphere emphasized distributional considerations. In line with \Cref{Proposition-Distributional,Prop-MajNegcor}, our analysis suggests that the political debate was ripe for the kind of zero-sum thinking that initially doomed the EESA bill to fail. Our analysis also speaks to why, following the subsequent market crash, the EESA bill passed on its second vote; that crash produced clear information about the aggregate benefits of stabilizing the financial system, rendering the  collective choice problem more advantageously correlated during the second vote.\footnote{Necessarily, our analysis abstracts from many factors, some of which increase the scope for zero-sum thinking. We draw attention to how some actors who were privy to the details of the EESA were closely connected to the financial industry. \cite*{mian2010political} also document that politicians supporting the EESA appear responsive to their constituents as well as to lobbying efforts and contributions from the financial industry. Allowing for those who gain from a policy to obtain better information would make a less informed voter even more distrustful than in our model.}

Another important case is the Brexit referendum of 2016. As discussed in the introduction, most estimates both before and after Brexit have indicated that being part of the EU benefits the UK economy, both workers and consumers. However, during the referendum, many details of the ``Leave'' policy were left unspecified, making its aggregate and distributional ramifications uncertain. Nevertheless, most accounts of Brexit paint its support as coming from a distrust of the London elite and the EU as well as the increased pressure on public services that came from waves of immigration; see, for example, \cite{olivas2019understanding}. In this setting, information was scarce and the debate largely focused on distributional considerations, both in terms of the contrast between elites and non-elites as well as the ``zero-sum game'' in public services that could be allocated to ``natives'' versus ``foreigners.'' 

More broadly, we view elements of our analysis to offer a plausible account for how economic anxiety fuels the demand for populist policies. In their survey, \cite{guriev2022} note several instances and suggest that ``identifying the exact mechanisms is still an open question.''\footnote{A notable example in the US context is the economic and political response to the China shock, which resulted in a significant drop in manufacturing employment and wages \citep{autor2016china}; \cite{autor2020importing} document how this shock has contributed to political polarization and much stronger support for politicians who run on protectionist platforms. Outside the US, \cite{dal2023economic} note how support for the populist radical-right Sweden-Democrats came from marginalized voters who were far more vulnerable to losing their jobs to immigrants.}
Our analysis offers an informational mechanism: suppose that voters do not know if losers will be compensated following greater trade or immigration liberalization, and look to the past to form these beliefs. Were there significant transfers in the past, then voters might see their interests as aligned; however, the absence of past transfers might lead them to worry more about the crowding-out effect, in line with \Cref{Theorem-ComparativeStatics}. Moreover, in line with \Cref{Prop-MajNegcor}, voters that are more vulnerable to trade and immigration have more reason to anticipate being  crowded out and may therefore be more inclined towards protectionist and populist policies. 

\section{Conclusion}\label{Section-Conclusion}

Our work describes a strategic mechanism for zero-sum thinking in elections, even if the policy choice is not zero sum. We find that voters may choose policies that do not match their collective preferences and information, and that equilibria may be fragile to asymmetric information. 

The central logic resembles \citeauthor{akerlof1970market}'s Lemons Problem: a voter votes against policies that command the support of others out of fear of what their support means for her. This line of thinking resonates with political rhetoric, particularly in the context of populist movements that suggest that the interests of the elites are fundamentally misaligned with those of the common voter. Although we model the equilibrium logic as applying at an individual level, one could have done so at a group-level, in the spirit of ethical voter models \citep{coate2004group,feddersen2006theory}.\footnote{Ethical voter models typically address costly voting; we consider the same calculus here but with costless voting. Say each voter of our model is a stand-in for workers in a different sector, and members of each group vote in the group's interest. Identical results then ensue if the collective choice problem is adversely correlated across these sectors.} We also see this logic as potentially manifesting in other political contexts, such as legislative action and lobbying, where the interested parties do not all expect to gain but have to act jointly to select a policy. Each player may then worry about being crowded out from gains when she conditions on others supporting a policy.

While our analysis tackles several important questions, it leaves others unanswered. We abstract from costly turnout decisions and it may be useful to see how endogenous participation affects collective choice. Equally, it would be interesting to see how parties strategically choose policies to exploit polarization.\footnote{This has been a theme of recent work, for example, \cite{buisseret2022polarization}, \cite{dziuda}, and \cite{levy/razin}.} One may also envision the design of information structures to capitalize on zero-sum thinking.
We hope to address these questions in future work.

\begin{singlespace}
    \addcontentsline{toc}{section}{References}
    \bibliographystyle{ecta}
    \bibliography{ams}

\begin{thebibliography}{53}
\newcommand{\enquote}[1]{``#1''}
\expandafter\ifx\csname natexlab\endcsname\relax\def\natexlab#1{#1}\fi

\bibitem[\protect\citeauthoryear{Acemoglu, Egorov, and Sonin}{Acemoglu et~al.}{2013}]{acemoglu2013political}
\textsc{Acemoglu, D., G.~Egorov, and K.~Sonin} (2013): \enquote{A Political Theory of Populism,} \emph{Quarterly Journal of Economics}, 128, 771--805.

\bibitem[\protect\citeauthoryear{Acharya}{Acharya}{2016}]{acharya2014equilibrium}
\textsc{Acharya, A.} (2016): \enquote{Information Aggregation Failure in a Model of Social Mobility,} \emph{Games and Economic Behavior}, 100.

\bibitem[\protect\citeauthoryear{Akerlof}{Akerlof}{1970}]{akerlof1970market}
\textsc{Akerlof, G.~A.} (1970): \enquote{The Market for ''Lemons": Quality Uncertainty and the Market Mechanism,} \emph{Quarterly Journal of Economics}, 84, 488--500.

\bibitem[\protect\citeauthoryear{Ali, Haghpanah, Lin, and Siegel}{Ali et~al.}{2022}]{ali2022sell}
\textsc{Ali, S.~N., N.~Haghpanah, X.~Lin, and R.~Siegel} (2022): \enquote{How to Sell Hard Information,} \emph{The Quarterly Journal of Economics}, 137, 619--678.

\bibitem[\protect\citeauthoryear{Ali, Mihm, Siga, and Tergiman}{Ali et~al.}{2021}]{ali2021adverse}
\textsc{Ali, S.~N., M.~Mihm, L.~Siga, and C.~Tergiman} (2021): \enquote{Adverse and Advantageous Selection in the Laboratory,} \emph{American Economic Review}, 111, 2152--78.

\bibitem[\protect\citeauthoryear{Austen-Smith and Banks}{Austen-Smith and Banks}{1996}]{austen1996information}
\textsc{Austen-Smith, D. and J.~S. Banks} (1996): \enquote{Information Aggregation, Rationality, and the Condorcet Jury Theorem,} \emph{American Political Science Review}, 90, 34--45.

\bibitem[\protect\citeauthoryear{Autor, Beck, Dorn, and Hanson}{Autor et~al.}{2024}]{autor2024help}
\textsc{Autor, D., A.~Beck, D.~Dorn, and G.~H. Hanson} (2024): \enquote{Help for the Heartland? The Employment and Electoral Effects of the Trump Tariffs in the United States,} Working Paper.

\bibitem[\protect\citeauthoryear{Autor, Dorn, Hanson, and Majlesi}{Autor et~al.}{2020}]{autor2020importing}
\textsc{Autor, D., D.~Dorn, G.~Hanson, and K.~Majlesi} (2020): \enquote{Importing political polarization? The electoral consequences of rising trade exposure,} \emph{American Economic Review}, 110, 3139--3183.

\bibitem[\protect\citeauthoryear{Autor, Dorn, and Hanson}{Autor et~al.}{2016}]{autor2016china}
\textsc{Autor, D.~H., D.~Dorn, and G.~H. Hanson} (2016): \enquote{The China Shock: Learning from Labor Market Adjustment to Large Changes in Trade,} \emph{Annual Review of Economics}, 8, 205--240.

\bibitem[\protect\citeauthoryear{Barelli, Bhattacharya, and Siga}{Barelli et~al.}{2022}]{barelli2022full}
\textsc{Barelli, P., S.~Bhattacharya, and L.~Siga} (2022): \enquote{Full Information Equivalence in Large Elections,} \emph{Econometrica}, 90, 2161--2185.

\bibitem[\protect\citeauthoryear{Bergeron, Carvalho, Henrich, Nunn, and Weigel}{Bergeron et~al.}{2024}]{carvalho2023zero}
\textsc{Bergeron, A., J.-P. Carvalho, J.~Henrich, N.~Nunn, and J.~Weigel} (2024): \enquote{Zero-Sum Thinking, the Evolution of Effort-Suppressing Beliefs, and Economic Development,} Working Paper.

\bibitem[\protect\citeauthoryear{Bhattacharya}{Bhattacharya}{2013}]{bhattacharya2013preference}
\textsc{Bhattacharya, S.} (2013): \enquote{Preference monotonicity and information aggregation in elections,} \emph{Econometrica}, 81, 1229--1247.

\bibitem[\protect\citeauthoryear{Bonomi, Gennaioli, and Tabellini}{Bonomi et~al.}{2021}]{bonomi2021identity}
\textsc{Bonomi, G., N.~Gennaioli, and G.~Tabellini} (2021): \enquote{Identity, beliefs, and political conflict,} \emph{The Quarterly Journal of Economics}, 136, 2371--2411.

\bibitem[\protect\citeauthoryear{Bueno De~Mesquita and Dziuda}{Bueno De~Mesquita and Dziuda}{2022}]{dziuda}
\textsc{Bueno De~Mesquita, E. and W.~Dziuda} (2022): \enquote{Partisan Traps,} Working Paper.

\bibitem[\protect\citeauthoryear{Buisseret and Van~Weelden}{Buisseret and Van~Weelden}{2022}]{buisseret2022polarization}
\textsc{Buisseret, P. and R.~Van~Weelden} (2022): \enquote{Polarization, valence, and policy competition,} \emph{American Economic Review: Insights}, 4, 341--352.

\bibitem[\protect\citeauthoryear{Carlsson and Van~Damme}{Carlsson and Van~Damme}{1993}]{carlsson1993global}
\textsc{Carlsson, H. and E.~Van~Damme} (1993): \enquote{Global games and equilibrium selection,} \emph{Econometrica: Journal of the Econometric Society}, 989--1018.

\bibitem[\protect\citeauthoryear{Chinoy, Nunn, Sequeira, and Stantcheva}{Chinoy et~al.}{2024}]{chinoy2022zero}
\textsc{Chinoy, S., N.~Nunn, S.~Sequeira, and S.~Stantcheva} (2024): \enquote{Zero-Sum Thinking and the Roots of US Political Divides,} Working Paper.

\bibitem[\protect\citeauthoryear{Coate and Conlin}{Coate and Conlin}{2004}]{coate2004group}
\textsc{Coate, S. and M.~Conlin} (2004): \enquote{A group rule--utilitarian approach to voter turnout: theory and evidence,} \emph{American Economic Review}, 94, 1476--1504.

\bibitem[\protect\citeauthoryear{Dal~B{\'o}, Finan, Folke, Persson, and Rickne}{Dal~B{\'o} et~al.}{2023}]{dal2023economic}
\textsc{Dal~B{\'o}, E., F.~Finan, O.~Folke, T.~Persson, and J.~Rickne} (2023): \enquote{{Economic and Social Outsiders but Political Insiders: Sweden's Populist Radical Right},} \emph{The Review of Economic Studies}, 90, 675--706.

\bibitem[\protect\citeauthoryear{Davidai and Ongis}{Davidai and Ongis}{2019}]{davidai2019politics}
\textsc{Davidai, S. and M.~Ongis} (2019): \enquote{The politics of zero-sum thinking: The relationship between political ideology and the belief that life is a zero-sum game,} \emph{Science Advances}, 5, eaay3761.

\bibitem[\protect\citeauthoryear{Ekmekci and Lauermann}{Ekmekci and Lauermann}{2020}]{ekmekci2020manipulated}
\textsc{Ekmekci, M. and S.~Lauermann} (2020): \enquote{Manipulated electorates and information aggregation,} \emph{The Review of Economic Studies}, 87, 997--1033.

\bibitem[\protect\citeauthoryear{Eliaz and Spiegler}{Eliaz and Spiegler}{2020}]{eliaz2020model}
\textsc{Eliaz, K. and R.~Spiegler} (2020): \enquote{A model of competing narratives,} \emph{American Economic Review}, 110, 3786--3816.

\bibitem[\protect\citeauthoryear{Fajgelbaum and Khandelwal}{Fajgelbaum and Khandelwal}{2022}]{fajgelbaum2022economic}
\textsc{Fajgelbaum, P.~D. and A.~K. Khandelwal} (2022): \enquote{The economic impacts of the US--China trade war,} \emph{Annual Review of Economics}, 14, 205--228.

\bibitem[\protect\citeauthoryear{Feddersen and Pesendorfer}{Feddersen and Pesendorfer}{1996}]{feddersen/pesendorfer:96}
\textsc{Feddersen, T. and W.~Pesendorfer} (1996): \enquote{The Swing Voter's Curse,} \emph{American Economic Review}, 86, 408--424.

\bibitem[\protect\citeauthoryear{Feddersen and Pesendorfer}{Feddersen and Pesendorfer}{1997}]{feddersen/pesendorfer:97}
---\hspace{-.1pt}---\hspace{-.1pt}--- (1997): \enquote{Voting Behavior and Information Aggregation in Elections with Private Information,} \emph{Econometrica}, 65, 1029--1058.

\bibitem[\protect\citeauthoryear{Feddersen and Pesendorfer}{Feddersen and Pesendorfer}{1998}]{feddersen1998convicting}
---\hspace{-.1pt}---\hspace{-.1pt}--- (1998): \enquote{Convicting the innocent: The inferiority of unanimous jury verdicts under strategic voting,} \emph{American Political science review}, 92, 23--35.

\bibitem[\protect\citeauthoryear{Feddersen and Sandroni}{Feddersen and Sandroni}{2006}]{feddersen2006theory}
\textsc{Feddersen, T. and A.~Sandroni} (2006): \enquote{A theory of participation in elections,} \emph{American Economic Review}, 96, 1271--1282.

\bibitem[\protect\citeauthoryear{Fernandez and Rodrik}{Fernandez and Rodrik}{1991}]{fernandez1991resistance}
\textsc{Fernandez, R. and D.~Rodrik} (1991): \enquote{Resistance to reform: Status quo bias in the presence of individual-specific uncertainty,} \emph{American Economic Review}, 81, 1146--1155.

\bibitem[\protect\citeauthoryear{Guriev and Papaioannou}{Guriev and Papaioannou}{2022}]{guriev2022}
\textsc{Guriev, S. and E.~Papaioannou} (2022): \enquote{The Political Economy of Populism,} \emph{Journal of Economic Literature}, 60, 753--832.

\bibitem[\protect\citeauthoryear{Halac, Lipnowski, and Rappoport}{Halac et~al.}{2021}]{halac2021rank}
\textsc{Halac, M., E.~Lipnowski, and D.~Rappoport} (2021): \enquote{Rank uncertainty in organizations,} \emph{American Economic Review}, 111, 757--786.

\bibitem[\protect\citeauthoryear{Inostroza and Pavan}{Inostroza and Pavan}{2023}]{inostroza2023adversarial}
\textsc{Inostroza, N. and A.~Pavan} (2023): \enquote{Adversarial coordination and public information design,} Working paper.

\bibitem[\protect\citeauthoryear{Iyengar, Lelkes, Levendusky, Malhotra, and Westwood}{Iyengar et~al.}{2019}]{iyengar2019origins}
\textsc{Iyengar, S., Y.~Lelkes, M.~Levendusky, N.~Malhotra, and S.~J. Westwood} (2019): \enquote{The origins and consequences of affective polarization in the United States,} \emph{Annual review of political science}, 22, 129--146.

\bibitem[\protect\citeauthoryear{Kajii and Morris}{Kajii and Morris}{1997}]{kajii1997robustness}
\textsc{Kajii, A. and S.~Morris} (1997): \enquote{The robustness of equilibria to incomplete information,} \emph{Econometrica}, 1283--1309.

\bibitem[\protect\citeauthoryear{Kim and Fey}{Kim and Fey}{2007}]{kim2007swing}
\textsc{Kim, J. and M.~Fey} (2007): \enquote{The swing voter's curse with adversarial preferences,} \emph{Journal of Economic Theory}, 135, 236--252.

\bibitem[\protect\citeauthoryear{Levendusky and Malhotra}{Levendusky and Malhotra}{2016}]{levendusky2016mis}
\textsc{Levendusky, M.~S. and N.~Malhotra} (2016): \enquote{(Mis) perceptions of partisan polarization in the American public,} \emph{Public Opinion Quarterly}, 80, 378--391.

\bibitem[\protect\citeauthoryear{Levy and Razin}{Levy and Razin}{2022}]{levy/razin}
\textsc{Levy, G. and R.~Razin} (2022): \enquote{Political social-learning: Short-term memory and cycles of polarization,} Working Paper.

\bibitem[\protect\citeauthoryear{Levy, Razin, and Young}{Levy et~al.}{2022}]{levy2022misspecified}
\textsc{Levy, G., R.~Razin, and A.~Young} (2022): \enquote{Misspecified politics and the recurrence of populism,} \emph{American Economic Review}, 112, 928--962.

\bibitem[\protect\citeauthoryear{Martin and Yurukoglu}{Martin and Yurukoglu}{2017}]{martin2017bias}
\textsc{Martin, G.~J. and A.~Yurukoglu} (2017): \enquote{Bias in cable news: Persuasion and polarization,} \emph{American Economic Review}, 107, 2565--99.

\bibitem[\protect\citeauthoryear{Mathevet, Perego, and Taneva}{Mathevet et~al.}{2020}]{mathevet/perego/taneva}
\textsc{Mathevet, L., J.~Perego, and I.~Taneva} (2020): \enquote{On Information Design in Games,} \emph{Journal of Political Economy}, 128, 1370--1404.

\bibitem[\protect\citeauthoryear{Meegan}{Meegan}{2010}]{meegan2010zero}
\textsc{Meegan, D.~V.} (2010): \enquote{Zero-sum bias: Perceived competition despite unlimited resources,} \emph{Frontiers in psychology}, 1, 191.

\bibitem[\protect\citeauthoryear{Mian, Sufi, and Trebbi}{Mian et~al.}{2010}]{mian2010political}
\textsc{Mian, A., A.~Sufi, and F.~Trebbi} (2010): \enquote{The political economy of the US mortgage default crisis,} \emph{American Economic Review}, 100, 1967--1998.

\bibitem[\protect\citeauthoryear{Morris and Shin}{Morris and Shin}{2012}]{morris2012contagious}
\textsc{Morris, S. and H.~S. Shin} (2012): \enquote{Contagious adverse selection,} \emph{American Economic Journal: Macroeconomics}, 4, 1--21.

\bibitem[\protect\citeauthoryear{Mudde}{Mudde}{2004}]{mudde2004populist}
\textsc{Mudde, C.} (2004): \enquote{The populist zeitgeist,} \emph{Government and opposition}, 39, 541--563.

\bibitem[\protect\citeauthoryear{Myerson}{Myerson}{1998}]{myerson1998population}
\textsc{Myerson, R.~B.} (1998): \enquote{Population uncertainty and Poisson games,} \emph{International Journal of Game Theory}, 27, 375--392.

\bibitem[\protect\citeauthoryear{Osuna, De~Lyon, Gartzou-Katsouyanni, Bulat, Kiefel, Bolet, Jablonowski, and Kaldor}{Osuna et~al.}{2019}]{olivas2019understanding}
\textsc{Osuna, J. J.~O., J.~M.~C. De~Lyon, K.~Gartzou-Katsouyanni, A.~Bulat, M.~Kiefel, D.~Bolet, K.~Jablonowski, and M.~Kaldor} (2019): \enquote{Understanding Brexit at a local level: causes of discontent and asymmetric impacts,} London School of Economics and Political Science.

\bibitem[\protect\citeauthoryear{P{\'a}stor and Veronesi}{P{\'a}stor and Veronesi}{2021}]{pastor2021inequality}
\textsc{P{\'a}stor, L. and P.~Veronesi} (2021): \enquote{Inequality aversion, populism, and the backlash against globalization,} \emph{The Journal of Finance}, 76, 2857--2906.

\bibitem[\protect\citeauthoryear{Perego and Yuksel}{Perego and Yuksel}{2022}]{perego2022media}
\textsc{Perego, J. and S.~Yuksel} (2022): \enquote{Media competition and social disagreement,} \emph{Econometrica}, 90, 223--265.

\bibitem[\protect\citeauthoryear{Petri and Plummer}{Petri and Plummer}{2016}]{petri2016economic}
\textsc{Petri, P.~A. and M.~G. Plummer} (2016): \enquote{The economic effects of the Trans-Pacific Partnership: New estimates,} Working Paper.

\bibitem[\protect\citeauthoryear{Razin}{Razin}{2003}]{razin2003signaling}
\textsc{Razin, R.} (2003): \enquote{Signaling and election motivations in a voting model with common values and responsive candidates,} \emph{Econometrica}, 71, 1083--1119.

\bibitem[\protect\citeauthoryear{R{\'o}{\.z}ycka-Tran, Boski, and Wojciszke}{R{\'o}{\.z}ycka-Tran et~al.}{2015}]{rozycka2015belief}
\textsc{R{\'o}{\.z}ycka-Tran, J., P.~Boski, and B.~Wojciszke} (2015): \enquote{Belief in a Zero-Sum Game as a Social Axiom: A 37-Nation Study,} \emph{Journal of Cross-Cultural Psychology}, 46, 525--548.

\bibitem[\protect\citeauthoryear{Sapienza and Zingales}{Sapienza and Zingales}{2013}]{SapienzaZingales}
\textsc{Sapienza, P. and L.~Zingales} (2013): \enquote{Economic Experts versus Average Americans,} \emph{American Economic Review}, 103, 636--42.

\bibitem[\protect\citeauthoryear{Szeidl and Szucs}{Szeidl and Szucs}{2024}]{szeidl2024}
\textsc{Szeidl, A. and F.~Szucs} (2024): \enquote{The Political Economy of Alternative Realities,} Working Paper.

\bibitem[\protect\citeauthoryear{Yuksel}{Yuksel}{2022}]{yuksel2022specialized}
\textsc{Yuksel, S.} (2022): \enquote{Specialized learning and political polarization,} \emph{International Economic Review}, 63, 457--474.

\end{thebibliography}
\end{singlespace} 


\appendix

\section{Appendix}

The appendix has the following structure:
\begin{itemize}[noitemsep]
    \item \Cref{App-Preliminaries} includes preliminary results and notation, describes the decomposition detailed in \Cref{Section-Scarce}, and contains all proofs for \Cref{Section-Model}.
    \item \Cref{App-Section3} contains all proofs for \Cref{Section-Results}.
\end{itemize}
The Online Appendix contains all proofs for \Cref{Section-UnpackingCorrelation,Section-Extensions}. 

\subsection{Preliminaries}\label{App-Preliminaries}

We first provide a formal description of the voting game and establish equilibrium existence, and then describe how we analyze equilibria with scarce information.

Throughout, $\mathcal{C}=(\mathcal{N},\Omega,P)$ refers to a collective choice problem, $\mathcal{G}=\{s^k\in \mathcal{M}: V^d_i(S_i=s^k)>0\}$ is the set of signals that convey good news, and $\mathcal{B}=\mathcal{M}\setminus\mathcal{G}$ the set of signals that convey bad news. For $g\in \{0,...,n-1\}$ and $m\in \{g,...,n-1\}$, 
\[
Z(g,m)=P(G=g|M=m)V_i^d(G=g,M=m,S_i=s^0)
\]
is the expected payoff difference for an uninformed voter who learns that $m$ other voters received news $g$ of whom received good news, weighted by the probability that $g$ voters receive good news when $m$ receive news. Finally $s_0$ is the signal profile where all voters receive the uninformative signal.

\subsubsection{Strategies and Equilibrium}

\paragraph{Private information:} A strategy-profile is a mapping $\sigma:\mathcal{N}\times \mathcal{S} \to [0,1]$, where $\sigma(i,s_i)\equiv \sigma_i(s_i)$ represents the probability that voter $i\in \mathcal{N}$ votes for $\pg $ when she receives the signal $s_ i \in \mathcal{S}$. Let $\Sigma$ be the set of all strategy-profiles, and let
\[
\Sigma^*_i = \{\sigma \in \Sigma: \forall\,i\in \mathcal{N}, \sigma_i(s^k)=1 \; \forall s^k\in \mathcal{G}\; \text{and}\;  \sigma_i(s^k)=0 \; \forall s^k\in \mathcal{B}\}.
\]
The set $\Sigma^*$  differs from $\Sigma$ only by excluding strategy-profiles where informed voters vote against their own signals, which are weakly dominated strategy-profiles by  \Cref{Assumption-Information}.\footnote{A strategy-profile $\sigma \in \Sigma$ is weakly undominated if, for each voter $i\in \mathcal{N}$, there does not exists an alternative strategy $\sigma_i'$ such that voter $i$'s expected payoff from $(\sigma_i',\sigma_{-i}')$ is greater than equal to her expected payoff from $(\sigma_i,\sigma_{-i}')$ for all $\sigma_{-i}'\in \Sigma_{-i}$ and strictly greater for some $\sigma_{-i}'\in \Sigma_{-i}$.} For a strategy-profile $\sigma \in \Sigma^*$, we simplify notation by letting $\sigma_i\equiv \sigma_i(s^0)$. 

An action-profile is a mapping $a:\mathcal{N}\to \{0,1\}$, where $a(i)=1$ represents a vote for $\pg $ and $a(i)=0$ represents a vote for $\pb $. Let $\mathcal{A}$ be the set of all action-profiles. For $a\in \mathcal{A}$, $a^{-1}(1) = \{i\in \mathcal{N}: a(i)=1\}$ is the set of voters who vote for $\pg $. For any voter $i\in \mathcal{N}$ and $a\in \mathcal{A}$, let $a^{-1}_{-i}(1)=a^{-1}(1)-\{i\}$.

Given a collective choice problem $\mathcal{C}=(\mathcal{N},\Omega,P)$, we denote by $P_{\sigma}$ the probability distribution on $\mathcal{A}\times \Omega$ induced by strategy-profile $\sigma \in \Sigma$ and the primitive  distribution $P$ on $\Omega$, defined by
\begin{align*}
P_{\sigma}(a,\omega)\equiv P(\omega) \prod_{i\in a^{-1}(1)} \sigma_i\big(S_i(\omega)\big) \prod_{j\in a^{-1}(0)}\Big(1-\sigma_j\big(S_j(\omega)\big)\Big).
\end{align*}
For a voter $i$, we then denote by $\Pi_i(\mathcal{C},\sigma)$ the difference between the expected payoff when voter $i$ votes for $\pg $ and the expected payoff when she votes for $\pb $ conditional on her receiving the uninformative signal $s^0$, which equates to
\begin{align*}
\Pi_i(\mathcal{C},\sigma)&\equiv \sum_{\{\omega\in \Omega:P(\omega)>0\}} V_i^d(\omega)P_{\sigma}\left(\big|a^{-1}_{-i}(1)\big|=\tau \Big| \omega \right) P(\omega|S_i=s^0)
\end{align*}
since voter $i$ impacts the election outcome only for action-profiles with $|a^{-1}_{-i}(1)|=\tau$. 

Let $\mathcal{N}_{i}(g,m)$ be the collection of all $(N_0,N_1)$ such that $N_0,N_1\subseteq \mathcal{N}-\{i\}$ with  $N_0\cap N_1 =\emptyset$, $ |N_0|=\tau -(m-g)$, and $|N_1|=\tau-g$. Then,  when $\sigma \in \Sigma^*$, 
\begin{align*}
\Pi_i(\mathcal{C},\sigma)=\sum_{g=0}^{\tau}\sum_{m=g}^{\tau+g}p_i(\sigma|g,m)P(M=m|S_i=s_0)Z(g,m)
\end{align*}
where\footnote{We follow the standard convention that the product over terms in the empty set is $1$.}
\begin{align}\label{Eq-Piv}
p_i(\sigma|g,m)&\equiv P_{\sigma}\left(|a^{-1}_{-i}(1)|=\tau \, \Big| \, G=g,M=m,S_i=s^0\right) \nonumber \\
&=\binom{n-1}{m}^{-1}\sum_{(N_0,N_1)\in \mathcal{N}_{i}(g,m)} \prod_{j\in N_1}\sigma_j(s^0)\prod_{k\in N_0}(1-\sigma_k(s^0)).
\end{align}

We observe that a strategy-profile $\sigma \in \Sigma$ is an equilibrium (in weakly undominated strategies) if and only if $\sigma \in \Sigma^*$ and, for all $i \in \mathcal{N}$, $\Pi_i(\mathcal{C},\sigma)>0$ implies $\sigma_i(s^0)=1$ and $\Pi_i(\mathcal{C},\sigma)<0$ implies $\sigma_i(s^0)=0$.

\paragraph{Public information:} In the public information benchmark, a strategy-profile is a mapping $\phi:\mathcal{N}\times \mathcal{S}^n \to [0,1]$ where $\phi(i,s)\equiv \phi_i(s)$ is the probability that voter $i$ votes for $\pg $ when she observes the signal-profile $s\in \mathcal{S}^n$.

\subsubsection{Proof of \Cref{Theorem-Existence}}

\begin{proof}[\unskip\nopunct]
In the private information baseline, let $\sigma^{\alpha}\in \Sigma^*$ denote the symmetric strategy-profile where $\sigma_i^{\alpha}=\alpha \in [0,1]$ for all $i\in \mathcal{N}$. Then, for $g\in \{0,...,\tau\}$ and $m\in \{g,...,g+\tau\}$,
\[
p_i(\sigma^{\alpha}|g,m)=\begin{cases}
\mathbbm{1}[g=\tau] &\text{if} \; \alpha=0\\
\mathbbm{1}[m-g=\tau] &\text{if}\; \alpha=1\\
\binom{n-1-m}{\tau-g}\alpha^{\tau-g}(1-\alpha)^{\tau-(m-g)} &\text{if}\; \alpha \in (0,1)
\end{cases},
\]
Hence, $p_i(\sigma^{\alpha}|g,m)$ is continuous in $\alpha$, and so $\Pi_i(\mathcal{C},\sigma^{\alpha})$ is continuous in $\alpha$. If $\Pi_i(\mathcal{C},\sigma^{1})\ge 0$,  then $\sigma^1$ is an equilibrium; if $\Pi_i(\mathcal{C},\sigma^{0})\le 0$, then $\sigma^0$ is an equilibrium; otherwise, there exists $\alpha^*\in (0,1)$ such that $\Pi_i(\mathcal{C},\sigma^{\alpha^*})=0$ and so $\sigma^{\alpha^*}$ is an equilibrium.

In the public information benchmark, the only weakly undominated strategy-profile is $\phi^*$, where $\phi^*_i(s)=\mathbbm{1}[V_i^d(s)>0]$ for all $i\in \mathcal{N}$, which is the unique equilibrium.  
\end{proof}

\subsubsection{Scarce information} 

For our equilibrium analysis with scarce information, the following Lemma shows how the probability distribution $P$ in a collective choice problem can be decomposed into the pair $(P_{\mathcal{I}},\lambda)$. Given a signal profile $s\in \mathcal{S}^n$, let $I(s)=\{i\in \mathcal{N}:s_i\in \mathcal{M}\}$ be the set of voters who  receive an informative signal,  and $C(s)=\{s'\in \mathcal{M}^n: s_i'=s_i \; \forall \, i\in I(s)\}$.



\begin{lemma}\label{Lemma-Decomp}
Let $\mathcal{C}=(\mathcal{N},\Omega, P)$ satisfy \Cref{Assumption-Exchangeability,Assumption-Information}. Then, for any state  $\omega=(v,s)\in \Omega$, 
\[
P(\omega) = \lambda^{M(s)}(1-\lambda)^{n-{M(s)}}\sum_{s'\in C(s)}P_{\mathcal{I}}(v,s').
\] 
Moreover, if $\sigma\in \Sigma^*$, then in each term of $\Pi_i(\mathcal{C},\sigma)$, $P(M=m|S_i=s^0)$ depends only on $\lambda$, $p_i(\sigma|g,m)$ depends only on $\sigma$, and $Z(g,m)$ depend only on $P_{\mathcal{I}}$. 
\end{lemma}

\begin{proof}
Given a signal profile $s\in \mathcal{S}^n$, let $s_I=\{(v,s')\in \Omega :s_i'=s_i\;\forall \, i\in I(s)\}$ and $s_{-I}=\{(v,s')\in \Omega:s_i=s^0 \; \forall \, i\notin I(s)\}$. A state $(v,s)=\{V=v\}\cap s_I\cap s_{-I}$ and so, following our notation conventions, $P(v,s_I,s_{-I})= P(v,s)$.

Let $\omega=(v,s)\in \Omega$ with $m=|I(s)|$. Then, by \Cref{Assumption-Exchangeability,Assumption-Information},
\begin{align*}
P(v,s)&=P(v,s_I,s_{-I})=P(v,s_I)(1-\lambda)^{n-m}
=\frac{P(v,s_I,S\in \mathcal{M}^n)}{\lambda^{n-m}}(1-\lambda)^{n-m}\\
&=\frac{P(v,s_I|S \in \mathcal{M}^n)\lambda^n}{\lambda^{n-m}}(1-\lambda)^{n-m}
=\lambda^m(1-\lambda)^{n-m}\sum_{s'\in C(s)}P_{\mathcal{I}}(v,s').
\end{align*}

Moreover, by \Cref{Assumption-Exchangeability,Assumption-Information},
\[
P(M=m|S_i=s^0)=\binom{n-1}{m}\lambda^m(1-\lambda)^{n-1-m},
\]
which depends only on $\lambda$. From \Cref{Eq-Piv} it is immediate that $p_i(\sigma|g,m)$ depends only on $\sigma$ when $g\le \tau$ and $m\le \tau+g$, and $p_i(\sigma|g,m)=0$ otherwise. Finally, we show that $Z(g,m)$ depends only on $P_{\mathcal{I}}$. For $g\in \{0,...,n-1\}$ and $m\in \{g,...,n-1\}$, let  $\Omega(g,m)=\{(v,s) \in \Omega: G(s)=g, M(s)=m\}$ and $\Omega_i(g,m)=\{\omega \in \Omega(g,m):S_i=s_0\}$. Then, 
\begin{align*}
P(G=g|M=m,S_i=s^0)&=\frac{\sum_{(v,s)\in \Omega(g,m)}\lambda^m(1-\lambda)^{n-m}\sum_{s'\in C(s)}P_{\mathcal{I}}(v,s')}{\binom{n-1}{m}\lambda^m(1-\lambda)^{n-m}}
\end{align*}
which depends only on $P_{\mathcal{I}}$, and so
\begin{align*}
V_i^d(G=g,M=m,S_i=s^0)&=\sum_{(v,s)\in \Omega_i(g,m)}P(v,s|G=g,M=m)v_i^d\\
&=\sum_{(v,s)\in \Omega_i(g,m)}\frac{P(v,s)}{P(G=g,M=m,S_i=s^0)}v_i^d\\
&=\sum_{(v,s)\in \Omega_i(g,m)}\frac{\lambda^m(1-\lambda)^{n-m}\sum_{s'\in C(s)}P_{\mathcal{I}}(v,s')}{P(G=g|M=m,S_i=s^0)\binom{n-1}{m}\lambda^m(1-\lambda)^{n-m}}v_i^d,
\end{align*}
depends only on $\mathcal{P}_{\mathcal{I}}$.
\end{proof}

\subsubsection{Proof of \Cref{Proposition-Public}}

\begin{proof}[\unskip\nopunct]
Fix $\varepsilon \in (0,1)$ and let $\lambda_{\varepsilon}= 1-(1-\varepsilon)^{\frac{1}{n}}$. Now consider the equilibrium strategy profile $\phi^*$ from the proof of  \Cref{Theorem-Existence}. Since $V^d_i(s_0)>0$, $\sigma_i^*(s_0) = 1$ for all $i\in \mathcal{N}$, and so $\pg $ wins in this event. Hence, for all $\lambda \in (0,\lambda_{\varepsilon})$, the probability that $\pg $ wins is greater than $P(S=s_0)=(1-\lambda)^n>(1-\lambda_{\varepsilon})^n=(1-\varepsilon)$.
\end{proof}

\subsection{Proofs for Section 3}\label{App-Section3}

In Section 3, we fix $(\mathcal N,\Omega_{\mathcal I},P_{\mathcal I})$, and look at the equilibrium outcomes of the family of collective choice problems $\{(\mathcal{N},\Omega,\mathcal P_{\mathcal I},\lambda):\lambda\in (0,1)\}$ as $\lambda$ gets small. By \Cref{Lemma-Decomp}, the correlation structure is a property of $\mathcal P_{\mathcal I}$: if any collective choice problem in $\{(\mathcal{N},\Omega,\mathcal P_{\mathcal I},\lambda):\lambda\in (0,1)\}$ is adversely/advantageously correlated, then all collective choice problems in the class have the same correlation. 

Since $\mathcal{P}_{\mathcal{I}}$ is fixed, we write the expected payoff difference $\Pi_i(\mathcal{C},\sigma)$ simply as a function of $\lambda$ and $\sigma$. For $\kappa\in \{0,...,\tau\}$, \cref{Assumption-NonRedundancy} implies that $P(G=\kappa|M=\kappa,S_i=s^0)>0$ and $V^G(\kappa)\ne 0$; hence, $Z(\kappa,\kappa)\ne 0$ and $Z(\kappa,\kappa)>0 \iff V^G(\kappa)>0$.

\subsubsection{Proof of \Cref{Theorem-MainResult}}

\begin{proof}[\unskip\nopunct]
Suppose the collective choice problem is adversely correlated: $V^G(\kappa)<0$ for some $\kappa \in \{1,...,\tau\}$. For any $t\in \{0,...,n\}$, let $\sigma^t$ be the strategy-profile in $\Sigma^*$ where $\sigma^t_i=\mathbbm{1}[i\le t]$ for all $i\in \mathcal{N}$ (i.e., the first $t$ voters vote for $\pg $ when uninformed and the remaining voters vote for $\pb $ when uninformed). Let $\mathcal{N}_t=\{1,...,t\}$ and $\mathcal{N}_{t}^c=\{t+1,...,n\}$.

Now fix $\varepsilon \in (0,1)$ and let $\lambda_{\varepsilon}= 1-(1-\varepsilon)^{\frac{1}{n}}$.  If $t\le \tau$, $\pb $ wins in the event $\{S=s_0\}$ and so, for all $\lambda \in (0,\lambda_{\varepsilon})$, the probability that $\pb $ wins is greater than $(1-\varepsilon)$. We can therefore complete the proof by showing that there exists $\bar{\lambda}$ such that, for all $\lambda <\bar{\lambda}$, $\sigma^t$ is an equilibrium for some $t\le \tau$. We do this by considering the two cases where $V^G(\tau)<0$ and $V^G(\tau)>0$. 

\medskip

\noindent\emph{Case 1:} When $V^G(\tau)<0$, $\sigma^0$ is an equilibrium for $\lambda$ sufficiently small.

For any voter $i\in \mathcal{N}$, $p_i(\sigma^0|g,m)=\mathbbm{1}[g=\tau]$ because only voters with good signals vote for $\pg $. Hence,
\begin{align*}
\Pi_i(\lambda,\sigma^0)&=\sum_{m=\tau}^{\tau+g}\binom{n-1}{m}\lambda^m(1-\lambda)^{n-1-m}Z(g,m).
\end{align*}
Since $\lambda^{\tau}(1-\lambda)^{\tau}>0$, it follows that $\Pi_i(\lambda,\sigma^0)<0$ if and only if
\[
\sum_{m=\tau}^{\tau+g}\binom{n-1}{m}\left(\frac{\lambda}{1-\lambda}\right)^{m-\tau}Z(g,m)<0,
\]
where the lhs converges to $\binom{n-1}{\tau}Z(\tau,\tau)<0$ as $\lambda\to 0$. 

\medskip

\noindent\emph{Case 2:} When $V^G(\tau)>0$, there exists $\kappa \in \{1,...,\tau-1\}$ such that $V^G(\kappa)<0$ and $V^G(\kappa')>0$ for $\kappa' \in \{\kappa+1,...,\tau\}$, and $\sigma^{\tau-\kappa}$ is an equilibrium for $\lambda$ sufficiently small.

For any voter $i\in \mathcal{N}_{\tau-\kappa}$, $p_i(\sigma^{\tau-\kappa}|g,m)=0$ if $g< \tau-(\tau-\kappa-1)=\kappa+1$ because there only $\tau-\kappa-1$ other voters who vote for $\pg $ when uninformed. Hence, 
\begin{align*}
\Pi_i(\lambda,\sigma^{\tau-\kappa})&=\sum_{g=\kappa+1}^{\tau}\sum_{m=g}^{\tau+g}\binom{n-1}{m}\lambda^m(1-\lambda)^{n-1-m}p_i(\sigma^{\tau-\kappa}|g,m)Z(g,m)
\end{align*}
Since $\lambda^{\kappa+1}(1-\lambda)^{n-1-(\kappa+1)}>0$, it follows that $\Pi_i(\lambda,\sigma^{\tau-\kappa})>0$ if and only if
\[
\sum_{g=\kappa+1}^{\tau}\sum_{m=g}^{\tau+g}\binom{n-1}{m}\left(\frac{\lambda}{1-\lambda}\right)^{m-(\kappa+1)}p_i(\sigma^{\tau-\kappa}|g,m)Z(g,m)>0,
\]
where the lhs converges to $\binom{n-1}{\kappa+1}p_i(\sigma^{\tau-\kappa}|\kappa+1,\kappa+1)Z(\kappa+1,\kappa+1)$ as $\lambda \to 0$. 
Since $i$ is pivotal in the non-null event where all $\kappa+1$ of the voters in  $\mathcal{N}_{\tau-\kappa}^c$ receive good news, $p_i(\sigma^{\tau-\kappa}|\kappa+1,\kappa+1)>0$, and $Z(\kappa+1,\kappa+1)>0$ because $V^G(\kappa+1)>0$. Hence, $\Pi_i(\lambda, \sigma^{\tau-\kappa})>0$ for $\lambda$ sufficiently small. 

For any voter $i\in \mathcal{N}_{\tau-\kappa}^c$, $p_i(\sigma^{\tau-\kappa}|g,m)=0$ if $g< \tau-(\tau-\kappa)=\kappa$ because there $\tau-\kappa$ other voters who vote for $\pg $ when uninformed. Hence, 
\begin{align*}
\Pi_i(\lambda,\sigma^{\tau-\kappa})&=\sum_{g=\kappa}^{\tau}\sum_{m=g}^{\tau+g}\binom{n-1}{m}\lambda^m(1-\lambda)^{n-1-m}p_i(\sigma^{\tau-\kappa}|g,m)Z(g,m)
\end{align*}
Since $\lambda^{\kappa}(1-\lambda)^{n-1-(\kappa)}>0$, it follows that $\Pi_i(\lambda,\sigma^{\tau-\kappa})<0$ if and only if
\[
\sum_{g=\kappa}^{\tau}\sum_{m=g}^{\tau+g}\binom{n-1}{m}\left(\frac{\lambda}{1-\lambda}\right)^{m-\kappa}p_i(\sigma^{\tau-\kappa}|g,m)Z(g,m)<0,
\]
where the lhs converges to $p_i(\sigma^{\tau-\kappa}|\kappa,\kappa)Z(\kappa,\kappa)$ as $\lambda \to 0$. Since $i$ is pivotal in the non-null event where $\kappa$ voters in $\mathcal{N}_{\tau-\kappa}^c$ receive good news, $p_i(\sigma^{\tau-\kappa}|\kappa,\kappa)>0$, and $Z(\kappa,\kappa)<0$ because $V^G(\kappa)<0$. Hence, $\Pi_i(\lambda, \sigma^{\tau-\kappa})<0$ for $\lambda$ sufficiently small. 

\medskip

Together, the two cases show that, with adverse correlation, there exists some $\bar{\lambda}\in (0,1)$ such that, for all $\lambda \in (0,\bar{\lambda})$ there is an equilibrium in which a majority of voters vote for $\pb $ when they are uninformed. Hence, $\pb $ wins when all voters are uninformed and so, for $\lambda \le \min\{\bar{\lambda},\lambda_{\varepsilon}\}$, there is an equilibrium in which $\pb $ wins with probability exceeding $1-\varepsilon$.

\medskip

For the converse, suppose the informed collective choice problem is advantageously correlated: $V^G(\kappa)>0$ for all $\kappa \in \{1,...,\tau\}$. Fix some $\varepsilon\in (0,1)$, let $\sigma \in \Sigma^*$  and, without loss of generality, let $\sigma_i \ge \sigma_{i+1}$ for $i=1,...,n-1$. We show that, for the strategy profile $\sigma$ and $\lambda$ sufficiently small, either $\pg $ wins with probability exceeding $1-\varepsilon$ or $\sigma$ is not an equilibrium. Hence, for $\lambda$ sufficiently small, if $\sigma$ is an equilibrium, then $\pg $ wins with probability exceeding $1-\varepsilon$.  

Let $\delta \equiv (1-\varepsilon)^{\frac{1}{2(\tau+1)}} \in (0,1)$ and $\lambda_{\varepsilon}'=1-(1-\varepsilon)^\frac{1}{2n}$. If $\sigma_{\tau+1} \ge \delta$, then, in the event $\{S=s_0\}$ where all voters are uninformed, voters $i=1,...,\tau+1$ vote for $\pg $ with probability at least $\delta$. Therefore, for all $\lambda \in (0,\lambda_{\varepsilon}')$,
\[
P_{\sigma}(|a^{-1}(1)|\ge \tau+1)\ge \delta^{\tau+1}(1-\lambda)^n> (1-\varepsilon)^{\frac{\tau+1}{2(\tau+1)}}(1-\varepsilon)^\frac{n}{2n}=1-\varepsilon.
\]
To complete the proof, it sufficies to show that, if $\sigma_{\tau+1} <  \delta$, then there exists $\lambda_{\varepsilon}$ such that, for all $\lambda \in (0,\lambda_{\varepsilon})$, $\sigma$ is not an equilibrium.

Therefore, suppose $\sigma_{\tau+1} <  \delta$. We consider voter $n$, who when uninformed votes $\pg $ with probability less than $1$, and show this is not a best-response. For $g\in\{0,...,\tau\}$ and $m\in \{g,...,\tau+g\}$, it follows from \Cref{Eq-Piv} that
\[
\binom{n-1}{m}^{-1}(1-\delta)^{\tau-(m-g)}\prod_{j=1}^{\tau-g} \sigma_j  \le p_n(\sigma|g,m)\le \binom{n-1-m}{\tau-g} \prod_{j=1}^{\tau-g} \sigma_j,
\]
and, therefore, for $g\in\{0,...,\tau\}$,
\begin{align*}
&\sum_{m=g}^{\tau+g}p_n(\sigma|g,m)P(M=m|S_n=s^0)Z(g,m) \ge \\
& \qquad \lambda^g(1-\lambda)^{n-1-g}\prod_{j=1}^{\tau-g} \sigma_j \left( (1-\delta)^{\tau}Z(g,g) - \sum_{m=g+1}^{\tau+g} \left(\frac{\lambda}{1-\lambda}\right)^{m-g} \mathcal{M}(m,g)\big|Z(g,m)\big| \right)
\end{align*}
where $\mathcal{M}(m,g)$ is shorthand for the multinomial coefficient
\[
\mathcal{M}(m,g)\equiv \binom{n-1}{m,\tau-g,\tau-(m-g)}.
\]
Since $\delta\in (0,1)$ and $Z(g,g)>0$, there exists $\lambda_g \in (0,1)$ such that
\[
(1-\delta)^{\tau}Z(g,g) - \sum_{m=g+1}^{\tau+g} \left(\frac{\lambda}{1-\lambda}\right)^{m-g} \mathcal{M}(m,g)\big|Z(g,m)\big| >0
\]
for all $\lambda \in (0,\lambda_g)$. In particular, $\lambda_g$ depends only on $\mathcal{C}_{\mathcal{I}}$. Let $\lambda_{\varepsilon}=\min \{\lambda_g:g\in \{0,...,\tau\}\}$. Since $\prod_{j=1}^{\tau-g} \sigma_j>0$ when $g=\tau$, it follows that $\Pi_n(\lambda,\sigma)>0$ for all $\lambda \in (0,\lambda_{\varepsilon})$. But then $\sigma$ is not an equilibrium for $\lambda \in (0,\lambda_{\varepsilon})$ because $\sigma_n<1$. 
\end{proof}

\subsubsection{Proof of \Cref{Theorem-GoodEquilibrium}}

\begin{proof}[\unskip\nopunct]

We first observe that, for $\sigma\in \Sigma^*$, $\Pi_i(\lambda,\sigma)$ can equivalently be written in terms of number of other voters who receive bad news:
\begin{align*}
\Pi_i(\lambda,\sigma)=\sum_{b=0}^{\tau}\sum_{m=b}^{\tau+b}q_i(\sigma|b,m)\binom{n-1}{m}\lambda^m(1-\lambda)^{n-1-m}Z_B(b,m)
\end{align*}
where, for $b\in \{0,...,\tau\}$ and $m\in \{b,...,\tau+b\}$,  $q_i(\sigma|b,m)=p_i(\sigma|m-b,m)$ and $Z_B(b,m)=Z(m-b,m)$.

Now fix $\varepsilon \in (0,1)$ and let $\sigma^{\alpha}\in \Sigma^*$ be the symmetric strategy-profile from the proof of \Cref{Theorem-Existence}, where $\sigma_i\equiv \alpha \in [0,1]$ for all $i\in\mathcal{N}$. We consider three cases.

\medskip

\noindent\emph{Case 1:} If $Z_B(\tau,\tau)>0$, then there exists $\lambda_{\varepsilon}$ such that, for all $\lambda \in (0,\lambda_{\varepsilon})$, $\sigma^1$ is an equilibrium in which $\pg $ wins with probability exceeding $1-\varepsilon$.

Let $\lambda_{\varepsilon}= 1-(1-\varepsilon)^{\frac{1}{n}}$. For the strategy-profile $\sigma^1$, $\pg $ wins in the event $\{S=s_0\}$ and so, for all $\lambda \in (0,\lambda_{\varepsilon})$, the probability that $\pg $ wins is greater than $(1-\varepsilon)$. We can therefore complete the proof by showing that $\Pi_i(\lambda,\sigma^1)>0$ for $\lambda$ sufficiently small. 

From \Cref{Eq-Piv}, $q_i(\sigma^1|b,m)=\mathbbm{1}[\tau=b]$ and so
\begin{align*}
\Pi_i(\lambda,\sigma^1)&=\lambda^\tau (1-\lambda)^{\tau}\sum_{m=\tau}^{n-1}\binom{n-1}{m}\left(\frac{\lambda}{1-\lambda}\right)^{m-\tau}Z_B(b,m)
\end{align*}
Since $\lambda^\tau (1-\lambda)^{\tau}>0$, it follows that $\Pi_i(\lambda,\sigma^1)>0$ if and only if
\[
\sum_{m=\tau}^{n-1}\binom{n-1}{m}\left(\frac{\lambda}{1-\lambda}\right)^{m-\tau}Z_B(b,m)>0
\]
where the lhs converges to $\binom{n-1}{\tau}Z_B(\tau,\tau)>0$ as $\lambda\to 0$. 

\medskip

\noindent\emph{Case 2:} If $Z_B(\tau,\tau)<0$, there exists $\lambda_{\varepsilon}$ such that, for all $\lambda \in (0,\lambda_{\varepsilon})$, there is $\alpha \in (0,1)$ such that $\sigma^{\alpha}$ is an equilibrium in which $\pg $ wins with probability exceeding $1-\varepsilon$.

Let $\lambda_{\varepsilon}=  1-(1-\varepsilon)^{\frac{1}{2n}} = \bar{\alpha}$. Then for all $\lambda\in (0,\lambda_{\varepsilon})$ and $\alpha \in (\bar{\alpha},1]$, the probability that $\pg $ wins for the strategy-profile $\sigma^{\alpha}$ in the event $\{S=s_0\}$ exceeds $\alpha^n$, and so $\pg $ wins with probability exceeding 
\[
\alpha^{n}(1-\lambda)^n > \bar{\alpha}^n (1-\lambda_{\varepsilon})^n = 1-\varepsilon.
\]
Hence, it suffices to show that there exists $\bar{\lambda} \in (0,\lambda_{\varepsilon})$ such that, for all $\lambda \in (0,\bar{\lambda})$, there exists $\alpha_{\lambda}\in (\bar{\alpha},1)$ such that $\sigma^{\alpha_{\lambda}}$ is an equilibrium.

Analogous to the argument in case 1, $Z_B(\tau,\tau)<0$ implies that there exists $\lambda_1 \in (0,\lambda_{\varepsilon})$ such that, for all $\lambda \in (0,\lambda_1)$, $\Pi_i(\lambda,\sigma^1)<0$. 

Since $\bar{\alpha}\in (0,1)$, for $b\in \{0,...,\tau\}$ and $m\in \{b,...,\tau+b\}$,
\begin{align*}
q_i(\sigma^{\bar{\alpha}}|b,m)=\binom{n-1-m}{\tau-b}(1-\bar{\alpha})^{\tau-b}\bar{\alpha}^{\tau-(m-b)}>0
\end{align*}
and, hence, 
\begin{align*}
\Pi_i(\lambda,\sigma^{\bar{\alpha}})&=(1-\lambda)^{n-1}\sum_{b=0}^{\tau}\sum_{m=b}^{\tau+b}\left(\frac{\lambda}{1-\lambda}\right)^{m}
\binom{n-1}{m}q_i(\sigma^{\bar{\alpha}}|b,m)Z_B(b,m),
\end{align*}
which converges to $\binom{n-1}{\tau}(1-\bar{\alpha})^{\tau}\bar{\alpha}^{\tau}Z_B(0,0)$ as $\lambda\to 0$. Since $Z_B(0,0)>0$, there exists $\lambda_2 \in (0,\lambda_{\varepsilon})$ such that, for all $\lambda \in (0,\lambda_2)$, $\Pi_i(\lambda,\sigma^{\bar{\alpha}})>0$.

Now let $\bar{\lambda} = \min\{\lambda_1, \lambda_2,\lambda_{\varepsilon}\}$ and let $\lambda \in (0,\bar{\lambda})$. Then, $\Pi_i (\lambda,\sigma^1)<0<\Pi_i(\lambda,\sigma^{\bar{\alpha}})$ and, since $\Pi_i(\lambda,\sigma^{\alpha})$ is continuous in $\alpha$,  there exists $\alpha_{\lambda}\in (\bar{\alpha},1)$ such that $\Pi_i(\lambda,\sigma^{\alpha_{\lambda}})=0$, and so $\sigma^{\alpha_{\lambda}}$ is an equilibrium in which $\pg $ wins with probability exceeding $1-\varepsilon$.

\medskip

\noindent\emph{Case 3:} If $Z_B(\tau,\tau)=0$ then $P(B=\tau|M=m)=0$ for all $m\in \{\tau,...,n-1\}$. Since $p_i(\sigma^1|m-b,m)=0$ for all $b\in \{0,...,\tau-1\}$ and $m\in\{b,...,\tau\}$, it follows that $\Pi_i(\lambda,\sigma^1)=0$ for all $\lambda$, and so $\sigma^1$ is an equilibrium in which $\pg $ wins in the event $\{S=s_0\}$.  If $\lambda \le 1-(1-\varepsilon)^{\frac{1}{n}}$, then $P(S=s_0)\ge 1-\varepsilon$, and so $\pg $ wins in an equilibrium with probability exceeding $1-\varepsilon$.

\medskip

Together, the three cases show that, with advantageous correlation, there exists $\lambda_{\varepsilon}\in (0,1)$ such that, for all $\lambda\in (0,\lambda_{\varepsilon})$, there exists $\alpha\in (0,1]$ such that  $\sigma^\alpha$ is a symmetric equilibrium in which $\pg $ wins with probability exceeding $1-\varepsilon$. 
\end{proof}

\newpage

\section{Online Appendix}

The Online Appendix contains all arguments, notation, and proofs for \Cref{Section-UnpackingCorrelation,Section-Extensions}. 

\subsection{Proofs for \Cref{Section-UnpackingCorrelation}}\label{App-Section4}

For a binary collective choice problem $\mathcal{C}=(\mathcal{N},\Omega,P)$, we let $W_i=\{\omega\in \Omega: V_i^d(\omega)>0\}$ be the set of states in which voter $i$ is a winner, and $\mathcal{W}(\omega)=\{i\in \mathcal{N}:\omega \in W_i\}$ denote the set of winners in state $\omega$.

\subsubsection{Proof of \Cref{Theorem-ComparativeStatics}}

\begin{proof}[\unskip\nopunct]
Suppose $\mathcal{C}$ is binary collective choice problem in which signals are fully-informative. We first show that, for any $\kappa \in \{0,...,\tau\}$, 
\begin{align}\label{Eq-FIVG}
V^G(\kappa|P_W,v)&= \sum_{w=\kappa}^n \left(\left(\frac{w-\kappa}{n-\kappa}\right) v_W - \left(\frac{n-w}{n-\kappa}\right)v_L \right)\frac{\binom{w}{\kappa}P_W(w)}{\sum_{w'=\kappa}^{n} \binom{w'}{\kappa}P_W(w')}.
\end{align}

For $\kappa \in \{0,...,\tau\}$, $w\in \{\kappa+1,...,n\}$, and $i\in \mathcal{N}$,
\begin{align*}
P(G=\kappa|S_i=s^0,M=\kappa,W=w)&=\left(\frac{w}{n}\right) \frac{\binom{w-1}{\kappa}\binom{n-1-(w-1)}{0}}{\binom{n-1}{\kappa} }+\left(\frac{n-w}{n}\right) \frac{\binom{w}{\kappa}\binom{n-1-w}{0}}{\binom{n-1}{\kappa}} = \frac{\binom{w}{\kappa}}{\binom{n}{\kappa}}
\end{align*}
and
\begin{align*}
P(G=\kappa|S_i=s^0,M=\kappa,W=\kappa)=\frac{1}{\binom{n}{\kappa}}.
\end{align*}
Therefore, for any $\kappa \in \{0,...,\tau\}$ and $w\in \{\kappa,...,n\}$, and $i\in \mathcal{N}$,
\begin{align*}
P(W=w|S_i=s^0,G=M=\kappa)& = \frac{\binom{w}{\kappa} P_W(w)}{ \sum_{w'=\kappa}^n \binom{w'}{\kappa} P_W(w')} .
\end{align*}
 \Cref{Eq-FIVG} then follows by observing that, 
 \begin{align*}
 P(W_i|S_i=s^0,G=M=\kappa,W=w)=\frac{w-\kappa}{n-\kappa}.
 \end{align*}

Parts (a) and (b) follow because $V^G(\kappa)$ is increasing in $v_W$ for a fixed $v_L$ and $P_W$, and decreasing in $v_L$ for a fixed $v_W$ and $P_W$.  Part (c) follows because, for $w'>w$, $P'_W(w')P_W(w) \geq P'_W(w)P_W(w')$ implies that
\begin{align*}
\frac{\binom{w'}{\kappa} P'_W(w')}{ \sum_{w''=\kappa}^n \binom{w''}{\kappa} P'_W(w'')} 
\frac{\binom{w}{\kappa} P_W(w)}{ \sum_{w''=\kappa}^n \binom{w''}{\kappa} P_W(w'')} \ge 
\frac{\binom{w}{\kappa} P'_W(w)}{ \sum_{w''=\kappa}^n \binom{w''}{\kappa} P'_W(w'')} 
\frac{\binom{w'}{\kappa} P_W(w')}{ \sum_{w''=\kappa}^n \binom{w''}{\kappa} P_W(w'')},
\end{align*}
and so $P_W\succeq_{LR} P_W'$ implies $P_W(.|S_i=s^0,G=M=\kappa)\succeq_{LR} P_W'(.|S_i=s^0,G=M=\kappa)$. Since $P(W_i|S_i=s^0,G=M=\kappa,W=w)$ is increasing in $w$, $P_W(.|S_i=s^0,G=M=\kappa)\succeq_{LR} P_W'(.|S_i=s^0,G=M=\kappa)$ then implies that $V^G(\kappa|P_W,v_W,v_L)\ge V^G(\kappa|P_W',v_W,v_L)$.
\end{proof}

\subsubsection{Proof of \Cref{Proposition-LocalAdverseCorrelation}}

\begin{proof}[\unskip\nopunct]
For $n$-voter collective choice problem, let
\[
V_n(g,b)=\left(\frac{\ceil{qn}-g}{n-g-b}\right)v_w-\left(\frac{n-\ceil{qn}-b}{n-g-b}\right)v_l
\]
be the expected payoff difference for an uninformed voter who learns that $g$ other voters received good news and $b$ other voters received bad news. Hence, the $n$-voter collective choice problem has adverse correlation if there exists $g\in \{0,...,\frac{n-1}{2}\}$ such that $V_n(g,0)<0$. In particular, since $V_n(g,0)$ is strictly decreasing in $g$, there is adverse correlation if and only if $V_n(\frac{n-1}{2},0)<0$,  i.e., 
\[
\frac{\vl}{\vw}> \frac{\ceil{qn}-\frac{n-1}{2}}{n-\ceil{qn}}=\frac{\frac{\ceil{qn}}{n}-\frac{1}{2}+\frac{1}{2n}}{1-\frac{\ceil{qn}}{n}}.
\]
The term on the right-hand side is strictly greater than $\rho$, and converges to $\rho$ as $n\to \infty$. Hence, if $\vl/\vw\le \rho$, the $n$-voter collective choice problem is advantageously correlated for any population size $n$; if $\vl/\vw>\rho$, then there exists $N(q,\vl/\vw)$ such that $n$-voter collective choice problem has adverse correlation for all $n>N(q,\vl/\vw)$.
\end{proof}

\subsubsection{Proof of \Cref{Proposition-LargePopulation}}

\begin{proof}[\unskip\nopunct]
Given $\lambda\in(0,1)$ and a strategy-profile $\sigma_{n}$ of the $n$-voter collective choice problem, let $P_{\sigma_{n},\lambda}(g,b|piv,S_i=s_{0})$ be the probability that an uninformed voter $i$ attaches to there being $g$ informed winners and $b$ informed losers, conditional on that voter being pivotal (as long as pivotality is a non-null event). If pivotality is a non-null event, the expected payoff difference conditional on being pivotal is, therefore,
\begin{align*}
\Pi_0^i(\sigma_n,\lambda)= & \sum_{g=0}^{\frac{n-1}{2}}\sum_{b=0}^{n-\ceil{ qn} }V_n(g,b)P_{\sigma_{n},\lambda}(g,b|piv,S_i=s_{0}),
\end{align*}
where we can omit the $i$-superscript if $\sigma_n$ is symmetric.

Now suppose $\vl/\vw>\rho$ and fix some $\varepsilon \in (0,1)$. For the $n$-voter collective choice problem, denote by $\sigma_n^*$ the symmetric weakly undominated strategy profile where uninformed voters vote for the inferior policy. Then, an uninformed voter is pivotal if and only if $g=\frac{n-1}{2}$, which is a non-null event, and
\[
P_{\sigma_{n}^*,\lambda}(g,b|piv,S_i=s_{0})=\begin{cases}
\binom{n-\ceil{qn}}{b}\lambda^b (1-\lambda)^{n-\ceil{qn}-b} &\text{if}\; g=\frac{n-1}{2}\\
0 &\text{otherwise}
\end{cases}.
\]
Therefore,  
\[
\Pi_0(\sigma_n^*,\lambda)=\sum_{b=0}^{n-\ceil{qn}}V_n\left(\frac{n-1}{2},b\right)\binom{n-\ceil{qn}}{b}\lambda^b (1-\lambda)^{n-\ceil{qn}-b},
\]
which is strictly increasing in $\lambda$ since $V_n\left(\frac{n-1}{2},b\right)$ is strictly increasing in $b$. Moreover, by the weak law of large numbers  and the Portmanteau Theorem, 
\[
\lim_{n\to \infty} \Pi_0(\sigma_n^*,\lambda) = \left(\frac{q-\frac{1}{2}}{\frac{1}{2}-\lambda(1-q)}\right)v_w-\left(\frac{(1-q)(1-\lambda)}{\frac{1}{2}-\lambda(1-q)}\right)v_l,
\]
which is strictly increasing in $\lambda$ and strictly negative if   $\lambda < 1-(\vl/\vw)^{-1}\rho$. Hence, for any $\lambda < 1-(\vl/\vw)^{-1}\rho$, there exists $N^*$ such that for all $n>N^*$, $\Pi_0(\sigma^*_n,\lambda)<0$; hence,   $\sigma_n^*$ is a strict equilibrium. Moreover, under the strategy-profile $\sigma_n^*$ only informed winners vote for the optimal policy. The proportion of informed winners is strictly increasing in $\lambda$ and, by the weak law of large numbers, converges in probability to $\lambda q$ as $n\to \infty$. Since $\lambda^* q < \frac{1}{2}$, it follows that there exists $N_{\varepsilon}$ such that for all $n>N_{\varepsilon}$, the inferior policy wins with probability greater than $1-\varepsilon$ when voters follow the strategy-profile $\sigma_n^*$. Hence, for all $n>\max\{N^*,N_{\varepsilon}\}$, there is a strict equilibrium in which the inferior policy wins with probability at least $1-\varepsilon$. 

Conversely, suppose $\vl/\vw\le \rho$ and fix any population size $n$ and $\lambda\in (0,1)$. Since $V_n(g,b)$ is strictly decreasing in $g$ and strictly increasing in $b$, it follows that for any $g\in \{0,...,\frac{n-1}{2}\}$ and $b\in \{0,...,n-\ceil{qn}\}$,
\[
V_n(g,b)\ge \left(\frac{\frac{\ceil{qn}}{n}-\frac{1}{2}+\frac{1}{2n}}{\frac{1}{2} + \frac{1}{2n}}\right)v_w - \left(\frac{1-\frac{\ceil{qn}}{n}}{\frac{1}{2} + \frac{1}{2n}}\right)v_l> \left(\frac{q-\frac{1}{2}}{\frac{1}{2}+\frac{1}{2n}}\right)v_w - \left(\frac{1-q}{\frac{1}{2}+\frac{1}{2n}}\right)v_l>0.
\]

Now, for sake for contradiction, suppose there exists a weakly undominated equilibrium $\sigma^n$ in which the inferior policy wins with strictly positive probability. Let $\mathcal{N}^*=\{i\in \{1,...,n\}: \sigma^n_i < 1\}$ be the set of voters who vote for the inferior policy with strictly positive probability when they are uninformed, and let $n^*=|\mathcal{N}^*|$. Since the inferior policy wins with strictly positive probability, $n-\ceil{qn}+n^* \ge \frac{n+1}{2}$, while $n-\ceil{qn}\le \frac{n-1}{2}$. Hence, any of the voters in $\mathcal{N}^*$ can be pivotal with strictly positive probability when they are uninformed. However, since $V_n(g,b)>0$ for all $g\in \{0,...,\frac{n-1}{2}\}$ and $b\in \{0,...,n-\ceil{qn}\}$, it follows that $\Pi_0^i(\sigma_n,\lambda)>0$, which contradicts that $\sigma^n$ is an equilibrium in which $\sigma^n_i<1$. Hence, in any weakly undominated equilibrium, the optimal policy wins with probability $1$.
\end{proof}

\subsubsection{Proof of \Cref{Proposition-Aggregate}}

\begin{proof}[\unskip\nopunct]
Suppose $\mathcal{C}$ is a binary collective choice problem in which signals convey only aggregate news. For any signal profile $s\in \mathcal{S}^n$, it follows that 
\begin{align*}
V_i^d(s) &= \sum_{w=0}^n V_i^d(s,W=w)P(W=w|S=s)=\sum_{w=0}^n V_i^d(W=w)P(W=w|S=s),
\end{align*}
and therefore, $V_i^d(s)=V_j^d(s)$ for all $i,j\in \mathcal{N}$. Now consider a signal profile $s$ such that $s_i \in \mathcal{G}$ and $s_j \in \mathcal{M}$ for some voters $i\ne j$. Then $V_i^d(s_i)>0$ and so $V_i^d(s)>0$ by \Cref{Assumption-Information}(b). It follows that $V_j^d(s)>0$, and so $V_j^d(s_j)>0$ by \Cref{Assumption-Information}(b). Hence, $s_j \in \mathcal{G}$. Analogously, if $s_i \in \mathcal{B}$ and $s_j \in \mathcal{M}$, then $s_j \in \mathcal{B}$. 

Let $G(\kappa)=\{s\in \mathcal{M}^n: G(s)\ge \kappa\}$. By \Cref{Lemma-Decomp} and the preceding argument, for a voter who is uninformed, and learns that $\kappa \in \{1,...,\tau\}$ other voters received good news,
\begin{align*}
P(W_i|S_i=s^0,M=G=\kappa)&=\lambda^{\kappa}(1-\lambda)^{n-\kappa} \sum_{s\in G(\kappa)} P(W_i|s)P(s|s\in G(\kappa))\\
&=\lambda^{\kappa}(1-\lambda)^{n-\kappa} \sum_{s\in \mathcal{G}^n} P(W_i|s)P(s|s\in \mathcal{G}^n)\\
&=\lambda^{\kappa}(1-\lambda)^{n-\kappa} P(W_i|S\in \mathcal{G}^n)
\end{align*}
and, therefore, $V^{G}(\kappa)>0$. 
\end{proof}

\subsubsection{Proof of \Cref{Proposition-Distributional}}

\begin{proof}[\unskip\nopunct]
We say that binary collective choice problems $\mathcal{C}=(\mathcal{N},\Omega,P)$ and $\mathcal{C}'=(\mathcal{N},\Omega',P')$ are informationally equivalent if, for any $\mathcal{N}'\subseteq \mathcal{N}$ and $s \in \mathcal{S}^n$, $P(\mathcal{W}=\mathcal{N}',S=s)=P'(\mathcal{W}=\mathcal{N}', S=s)$; that is, the problems differ in terms of the payoff but not the information structure. We show that, when all news is distributional, for any binary collective choice problem, there is an informationally equivalent problem that has adverse correlation.

\medskip

Suppose $\mathcal{C}=(\mathcal{N},\Omega,P)$ is a binary collective choice problem in which signals convey only distributional news. Without loss of generality, let $P(W_i|S_i=s^{k})\ge P(W_i|S_i=s^{k+1})$ for $k=1,...,K-1$ (where it does not matter in the following how ties are broken). Let $\mathcal{G}'=\{k=1,...,K:P(W_i|S_i=s^K)\ge P(W_i)\}$ and $\mathcal{B}'=\mathcal{M}-\mathcal{G}'$. Since $P(B\ge 1)>0$, both $\mathcal{G}'$ and $\mathcal{B}'$ are non-empty, and $P(W_i|S_i=s^1)>P(W_i)$. For a signal profile $s\in \mathcal{S}^n$,  $G'(s)$ is the number of voters with a signal in $\mathcal{G}'$. 

We first show that, for any voter $h\in \mathcal{N}$, 
\begin{align}\label{Eq-Distributional}
P(W_h)>P(W_h|S_h=s^0,G'=M=1).
\end{align} 

Let $w\in \{0,...,n\}$ and $E$ be any event such that $E\cap W^{-1}(w)$ is non-null. Then,
\begin{align*}
\sum_{i=1}^n P(W_i| E, w)&= \sum_{i=1}^n \sum_{\omega \in \Omega(w)\cap E} P(W_i|E,w,\omega) P(\omega| E, w)\\
&=   \sum_{\omega \in \Omega(w)\cap E}  P(\omega| E, w) \sum_{i=1}^n P(W_i|E,w,\omega)\\
&= w \sum_{\omega \in \Omega(w)\cap E} P(\omega |E, w)= w 
\end{align*}
Therefore, for any voter $i\ne h$ and $w\in  \{0,...,n\}$ with $P(W=w)>0$, \Cref{Assumption-Exchangeability} implies that
\begin{align*}
\sum_{j=1}^n P(W_j | S_i\in \mathcal{G}',  M=1, w) &= P(W_i |S_i\in \mathcal{G}', M=1, w) + \sum_{j\ne i} P(W_j|S_i\in \mathcal{G}', M=1,w)\\
&= P(W_i |S_i\in \mathcal{G}', M=1, w) + (n-1) P(W_h|S_i\in \mathcal{G}', M=1,w)\\
&=\sum_{j=1}^n P(W_j | w) = P(W_i |w) + \sum_{j\ne i} P(W_j|w)\\
&= P(W_i |w) + (n-1) P(W_h|w).
\end{align*}
Since $P(W_i|w)<P(W_i|S_i\in \mathcal{G}',M=1,w)$, it follows that  $P(W_h|w) > P(W_h|S_i\in \mathcal{G}',M=1,w)$. Moreover, by \Cref{Assumption-Exchangeability},
\begin{align*}
P(W_h|S_k=s^0,G'=M=1,w) &= \sum_{j\ne h} P(W_h|S_j\in \mathcal{G}',M=1,w) P(S_j\in \mathcal{G}'|G'=M=1,w)\\
&= \frac{1}{n-1} \sum_{j\ne h} P(W_h|S_j \in \mathcal{G}',M=1,w)\\
&= P(W_h|S_i\in \mathcal{G}',M=1,w),
\end{align*}
and, therefore,  $P(W_h|w) < P(W_h|S_k=s^0,G'=M=1,w)$. Since signals convey only distributional information, 
\begin{align*}
P(W_h)&=\sum_{w=0}^n P(W_h|w)P(w)\\
&< \sum_{w=0}^n P(W_h|S_k=s^0,G'=M=1,w)P(w)\\
&= \sum_{w=0}^n P(W_h|S_k=s^0,G'=M=1,w)P(w|S_h=s^0,G'=M=1)\\
&=P(W_h|S_h=s^0,G'=M=1),
\end{align*}
and, therefore, $P(W_h)>P(W_h|S_k=s^0,G'=M=1)$. 

Now let $k^* = \min\{k=1,...,K: s^k \in \mathcal{B}'\}$ and let $P^*= \max\{P(W_i|S_i = s^{k^*}),P(W_h|S_k=s^0,G'=M=1)\}$. From the preceding argument, $P(W_h)>P^*$ and so there exists $(v_W',v_L')>>0$ such that 
\[
P(W_h)v_W'-(1-P(W_h))v_L' > 0 > P^*v_W'-(1-P^*)v_L'.
\]
The binary collective choice problem $\mathcal{C}'=(\mathcal{N},\Omega,P')$ uniquely defined by letting $P'(\mathcal{W}=\mathcal{N}',S=s)=P(\mathcal{W}=\mathcal{N}', S=s)$ for any $\mathcal{N}'\subseteq \mathcal{N}$ and $s \in \mathcal{S}$ is informationally equivalent to $\mathcal{C}$, and $\mathcal{C}'$ is adversely correlated because the set of good news signals for $\mathcal{C}'$ is exactly $\mathcal{G}'$. As a result, for any binary collective choice problem $\mathcal{C}$ in which signals convey only distributional information there exists an informationally equivalent collective choice problem with adverse correlation. 
\end{proof}

\subsection{Proofs for \Cref{Section-Extensions}}
\subsubsection{Symmetric Equilibria (\Cref{Section-Symmetry})}

\paragraph{Preliminaries:} Let $\sigma^{\alpha}$ be the symmetric strategy-profile defined in the proof of \Cref{Theorem-Existence}. Recall that, for $g\in \{0,...,\tau\}$ and $m\in \{g,...,g+\tau\}$,
\[
p_i(\sigma^{\alpha}|g,m)=\begin{cases}
\mathbbm{1}[g=\tau] &\text{if} \; \alpha=0\\
\mathbbm{1}[m-g=\tau] &\text{if}\; \alpha=1\\
\binom{n-1-m}{\tau-g}\alpha^{\tau-g}(1-\alpha)^{\tau-(m-g)} &\text{if}\; \alpha \in (0,1)
\end{cases},
\]
and so
\begin{align*}
\Pi_0(\sigma^1,\lambda)&=\sum_{g=0}^{\tau} \binom{n-1}{\tau+g} \lambda^{\tau+g}(1-\lambda)^{\tau-g}Z(g,\tau+g),\\
\Pi_0(\sigma^0,\lambda)&=\sum_{m=\tau}^{n-1} \binom{n-1}{m} \lambda^{m}(1-\lambda)^{n-1-m}Z(\tau,m),
\end{align*}
and, for $\alpha \in (0,1)$,
\begin{align*}
&\Pi_0(\sigma^{\alpha},\lambda)=\sum_{g=0}^{\tau}\sum_{m=g}^{\tau+g} \mathcal{M}(g,m) \lambda^{m}(1-\lambda)^{n-1-m}\alpha^{\tau-g}(1-\alpha)^{\tau+g-m}Z(g,m),\\
&=\alpha^{\tau}(1-\alpha)^{\tau} (1-\lambda)^{n-1} \sum_{g=0}^{\tau} \left(\frac{\lambda(1-\alpha)}{\alpha(1-\lambda)}\right)^g\sum_{m=g}^{\tau+g} \left(\frac{\lambda}{(1-\alpha)(1-\lambda)}\right)^{m-g}\mathcal{M}(g,m)Z(g,m)
\end{align*}
where $\mathcal{M}(g,m)$ is shorthand for the multinomial coefficient $\binom{n-1}{\tau-g,m,\tau+g-m}$.
 
\subsubsection*{Proof of \Cref{Theorem-Symmetric}}

\begin{proof}[\unskip\nopunct]
(1) Suppose $\mathcal{C}$ is strongly adversely correlated: $\sum_{\kappa=0}^{\tau} \theta^{\kappa} \binom{\tau}{\kappa} Z(\kappa,\kappa)<0$ for some $\theta \in \mathbbm{R}_{++}$, and fix some $\varepsilon\in (0,1)$. 

By Case 1 in the proof of \Cref{Theorem-MainResult}, if $Z(\tau,\tau)<0$, then there exists $\bar{\lambda}\in (0, 1-(1-\varepsilon)^{\frac{1}{n}})$ such that  $\sigma^0$ is a symmetric equilibrium in which $\pb $ wins with probability exceeding $1-\varepsilon$ for all $\lambda \in (0,\bar{\lambda})$.  Therefore, we can focus on the case $Z(\tau,\tau)>0$. In that case, since 
\[
\lim_{\lambda\to 0} \sum_{m=\tau}^{n-1} \binom{n-1}{m} \lambda^{m-\tau}(1-\lambda)^{n-1-m}Z(\tau,m) =\binom{n-1}{\tau} Z(\tau,\tau)
\]
it follows that $\Pi_0(\sigma^0,\lambda)>0$ for $\lambda>0$ sufficiently small.

Let $\bar{\alpha}\equiv1-(1-\varepsilon)^{\frac{1}{2(\tau+1)}}$. If $\alpha < \bar{\alpha}$ and  $\lambda<1-(1-\varepsilon)^{\frac{1}{2n}}$, then $\pb $ wins in strategy profile $\sigma^{\alpha}$ with probability exceeding
\[
(1-\bar{\alpha})^{\tau+1}(1-\lambda)^n > (1-\varepsilon)^{\frac{\tau+1}{2(\tau+1)}}(1-\varepsilon)^{\frac{n}{2n}}=1-\varepsilon
\] 
Since $\Pi_0(\sigma^0,\lambda)>0$ for $\lambda$ sufficiently small, it therefore sufficies to show that there exists $\bar{\lambda}\in (0,1)$ such that, for all $\lambda\in (0,\bar{\lambda})$, there is a $\alpha_{\lambda}\in (0,\bar{\alpha})$ such that $\Pi_0(\sigma^{\alpha_{\lambda}},\lambda)<0$.

For any $\lambda \in (0,1)$, let $\alpha_{\lambda}\equiv \frac{\lambda}{(1-\lambda)\theta+\lambda}$; hence, $\alpha_{\lambda}\in (0,1)$, is increasing in $\lambda$, and converges to $0$ as $\lambda\to 0$. Therefore,
\begin{align*}
&\lim_{\lambda \to 0} \sum_{g=0}^{\tau} \left(\frac{\lambda(1-\alpha_{\lambda})}{\alpha_{\lambda}(1-\lambda)}\right)^g\sum_{m=g}^{\tau+g} \left(\frac{\lambda}{(1-\alpha_{\lambda})(1-\lambda)}\right)^{m-g}\mathcal{M}(g,m)Z(g,m)\\
 &\qquad \qquad = \sum_{g=0}^{\tau} \theta^g\mathcal{M}(g,g)Z(g,g) = \binom{n-1}{\tau}  \Kappa(\theta) <0.
\end{align*}
Hence, there is $\bar{\lambda}\in (0,1)$ such that $\Pi_0(\sigma^{\alpha_{\lambda}},\lambda)<0$ for all $\lambda \in (0,\bar{\lambda})$.

\medskip

\noindent (2) Suppose $\mathcal{C}$ is weakly advantageously correlated: $\sum_{\kappa=0}^{\tau} \theta^{\kappa} \binom{\tau}{\kappa} Z^G(\kappa)>0$ for all $\theta \in \mathbbm{R}_{++}$. 

The correlation structure implies that $Z(\tau,\tau)>0$, and so there exists $\bar{\theta}$ such that $\sum_{\kappa=0}^{\tau} \theta^{\kappa} \binom{\tau}{\kappa} Z(\kappa,\kappa)>Z(0,0)$ for all $\theta>\bar{\theta}$. Since $\lim_{\theta\to 0} \sum_{\kappa=0}^{\tau} \theta^{\kappa} \binom{\tau}{\kappa} Z(\kappa,\kappa)=Z(0,0)$, it follows that $\Kappa$ attains a minimum on $[0,\bar{\theta}]$, which is strictly positive. As a result, there exists  $\delta>0$ such that $\sum_{\kappa=0}^{\tau} \theta^{\kappa} \binom{\tau}{\kappa} \left(Z^G(\kappa)-\delta \right)>0$ for all $\theta \in \mathbbm{R}_{++}$. 

Now fix some $\varepsilon\in (0,1)$, and let $\bar{\alpha}\equiv (1-\varepsilon)^{\frac{1}{2(\tau+1)}} \in (0,1)$. If $\alpha\in[ \bar{\alpha},1]$ and $\lambda < 1-(1-\varepsilon)^{\frac{1}{2n}}$, then $\pg $ wins with probability exceeding
\[
\alpha^{\tau+1}(1-\lambda)^n > (1-\varepsilon)^{\frac{\tau+1}{2(\tau+1)}}(1-\varepsilon)^{\frac{n}{2n}}=1-\varepsilon
\]
in the strategy profile $\sigma^{\alpha}$. 

For $g\in \{0,...,\tau\}$, let $\phi(g)\equiv  \mathcal{M}(g,g)Z(g,g)$ and, for $\alpha,\lambda \in (0,1)$, let
\begin{align*}
\phi(g,\alpha,\lambda)&\equiv \sum_{m=g+1}^{\tau+g} \left(\frac{\lambda}{(1-\alpha)(1-\lambda)}\right)^{m-g}\mathcal{M}(g,m)Z(g,m)
\end{align*}
so that
\begin{align*}
\Pi_0(\sigma^{\alpha},\lambda)&= \alpha^{\tau}(1-\alpha)^{\tau} (1-\lambda)^n \sum_{g=0}^{\tau} \left(\frac{\lambda(1-\alpha)}{\alpha(1-\lambda)}\right)^g \Big(\phi(g) + \phi(g,\alpha,\lambda)\Big)\\
&\ge  \alpha^{\tau}(1-\alpha)^{\tau} (1-\lambda)^n \sum_{g=0}^{\tau} \left(\frac{\lambda(1-\alpha)}{\alpha(1-\lambda)}\right)^g \Big(\phi(g) - |\phi(g,\alpha,\lambda)|\Big)
\end{align*}
For $g\in \{0,...,\tau\}$ and $\alpha \in (0,\bar{\alpha})$, $|\phi(g,\alpha,\lambda)|\le |\phi(g,\bar{\alpha},\lambda)|$, and so there exists $\bar{\lambda}\in (0,1)$ such that $|\phi(g,\bar{\alpha},\lambda)| \le \binom{n-1}{\tau}\delta  $ for all $\lambda \in (0,\bar{\lambda})$ and $g\in \{0,...,\tau\}$. Hence, for all $\lambda \in (0,\bar{\lambda})$, 
\begin{align*}
\Pi_0(\sigma^{\alpha},\lambda) &\ge \alpha^{\tau}(1-\alpha)^{\tau} (1-\lambda)^n \sum_{g=0}^{\tau} \left(\frac{\lambda(1-\alpha)}{\alpha(1-\lambda)}\right)^g\left(\phi(g)-\binom{n-1}{\tau}\delta\right)\\
&= \alpha^{\tau}(1-\alpha)^{\tau} (1-\lambda)^n \binom{n-1}{\tau} \sum_{\kappa=0}^{\tau} \left(\frac{\lambda(1-\alpha)}{\alpha(1-\lambda)}\right)^{\kappa}\binom{\tau}{\kappa}\left(Z(\kappa,\kappa)-\delta \right)>0,
\end{align*}
since $\frac{\lambda(1-\alpha)}{\alpha(1-\lambda)}\in \mathbbm{R}_{++}$. Therefore, for all $\lambda \in (0,\bar{\lambda})$, $\sigma^{\alpha}$ is not an equilibrium for any $\alpha \in (0,\bar{\alpha})$. Moreover, if $Z(\tau,\tau)>0$, there exists $\bar{\lambda}_0\in(0,1)$ such that $\Pi_0(\sigma^0,\lambda)>0$ for all $\lambda \in (0,\bar{\lambda}_0)$.

Hence, for $\lambda_{\varepsilon}=\min\{ 1-(1-\varepsilon)^{\frac{1}{2n}},\bar{\lambda},\bar{\lambda}_0\}$,  the preceding arguments show that, for all $\lambda \in (0,\lambda_{\varepsilon})$ and any $\alpha \in [0,1]$, either $\pg $ wins with probability exceeding $(1-\varepsilon)$ in the strategy profile $\sigma^{\alpha}$ or the strategy profile $\sigma^{\alpha}$ is not an equilibrium.
\end{proof}

\subsubsection{Population Uncertainty (\Cref{Section-Population})}

\paragraph{Preliminaries:} We first describe how we adapt our main assumptions from \Cref{Section-Model} to the setting with population uncertainty. As previously, let $\mathcal{M}=\{s^1,...,s^K\}$ be the set of informative signals for any population size.  For $\omega \in \Omega^n$, $V(\omega)$ and $S(\omega)$ are the payoff and signal profiles in state $\omega$, and $V_i^d(E,n)$ is the expected payoff difference between the ex-ante optimal and inferior policies for a voter $i$  who conditions on the population size $n$ and the event $E\subseteq \Omega^n$. We continue to define $V_i^d(E,n)\equiv 0$ when $E$ is a null-event and assume that $V_i^d(E,n)\ne 0$ otherwise.

\begin{assumption}\label{Assumption-Exchangeable-RP}
Voters are exchangeable for any population size $n\in \mathcal{Q}$: if $\omega,\omega'\in \Omega^n$ and $\omega$ permutes $\omega'$, then $P_n(\omega)=P_n(\omega')$.
\end{assumption}

Let $\pg _n$ be the optimal policy when voters learn only that the population size is $n$. By \cref{Assumption-Exchangeable-RP}, voters agree on $\pg _n$. We assume that $\pg _n$ does not depend on $n$.

\begin{assumption}
For all $n,n'\in \mathcal{N}$,  $\pg _n=\pg _{n'}$.
\end{assumption}

\begin{assumption}\label{Assumption-Information-RP}
There is an uninformative signal, and other signals are sufficient:
\begin{enumerate}[label=\emph{(\alph*)},ref=(\alph*),nolistsep]
\item \label{independence}
\emph{\textbf{Uninformative signal:}} For $n\in \mathcal{Q}$, $\omega\in \Omega^n$ with $S_i(\omega)=s^0$, 
\[
P_n(\omega)=P_n(V(\omega),S_{-i}(\omega))(1-\lambda).
\] 
for some $\lambda \in (0,1)$. 
\item \label{sufficiency}
\emph{\textbf{Informative signals:}} For $n\in \mathcal{Q}$ and $s_i \in \mathcal{M}$, $V_i^d(s_i,n)>0$ if and only if $V_i^d(s',n')>0$ for all $n'\in \mathcal{Q}$, $s'\in \mathcal{S}^{n'}$ such that $s_i'=s_i$.
\end{enumerate}
\end{assumption}

By \Cref{Assumption-Information-RP}, we can again classify informative signals as good or bad news.
We let $\tau_0\equiv \tau(n_0)$ and $P_0\equiv P_{n_0}$. 

\begin{assumption}\label{Assumption-New-RP}
$P_0(B\ge 1)>0$ and $P_0(G\ge \tau_0)>0$.
\end{assumption}

We denote the mean population size by $\mu$ and the CDF of $Q$ by $F$. As observed by \cite{myerson1998population}, being selected to participate in an election, leads a voter to update their beliefs about the size of the electorate. To perform this updating when $\mathcal{Q}$ may be countably infinite, we follow  \cite{myerson1998population} by first assuming $\bar{N}\in \mathcal{Q}$ players are pre-selected, each of whom is equally likely to be recruited as a voter. We then calculate voter $i$'s beliefs about the size of the electorate, conditional on the event $R_i$  that $i$ is a voter, and take the limit as $\bar{N}\to \infty$. Hence, 
\begin{align*}
Q(N=n|R_i) &\equiv \lim_{\bar{N}\to \infty} Q(N=n | R_i, N\le \bar{N} )\\
&=\lim_{\bar{N}\to \infty} \frac{Q(R_i|N=n,N\le\bar{N})Q(N=n|N\le \bar{N})}{\sum_{n'=n_0}^{\bar{N}}Q(R_i|N=n',N\le \bar{N})Q(N=n|N\le \bar{N})}\\
&= \lim_{\bar{N}\to \infty} \frac{\frac{n}{\bar{N}}\frac{\mathbbm{1}[n\le \bar{N}]Q(N=n)}{F(\bar{N})}}{\sum_{n'=n_0}^{\bar{N}}\frac{n'}{\bar{N}}\frac{Q(N=n')}{F(\bar{N})}}
= \frac{nQ(N=n)}{\mu}
\end{align*}

By \Cref{Assumption-Information-RP}(b), a voter who receives an informative signal has a unique undominated action. Given the population uncertainty, we focus on the set of symmetric undominated strategy profiles $\sigma^{\alpha}$, where voters with good signals vote for $\pg $, voters with bad signals vote for $\pb $, and voters with the signal $s^0$ independently vote for $\pg $ with probability $\alpha$ for some $\alpha \in [0,1]$. Adapting our previous notation, let $\Pi(\alpha,\lambda)$ be the expected payoff difference between a vote for $\pg $ and vote for $\pb $ for a voter who receives signal $s^0$ when $P(S_i\in \mathcal{M})=\lambda$ and other voters follow the strategy-profile $\sigma^{\alpha}$. A subscript $n$ means ``conditional on population size $n$," with a subscript $0$ for the case when $n=n_0$. By  \Cref{Assumption-Information-RP}(a), signal $s^0$ is not informative about the population size, payoff-profile or signal-profile of the other voters. Hence, for a voter $i$ and $m\in \{0,...,n-1\}$,
\[
P_n(M=m|S_i=s^0,N=n)=\binom{n-1}{m}\lambda^m(1-\lambda)^{n-1-m}.
\]
We formulate the following elementary property of absolutely convergent series for later reference.

\begin{lemma}\label{Lemma-S}
Let $a:\mathbb{N}^{2}\to\mathbb{R}$ such that, for all $t$, $\sum_{n=0}^{\infty}a(n,t)$ is absolutely convergent and, for all $n$, $a(n,t)$ converges monotonically to $0$. Then, $\lim_{t\to0}\sum_{n=0}^{\infty}a(n,t)=0$.
\end{lemma}

\begin{proof}
For $(n,t)\in\mathbb{N}^{2}$, let $a^{+}(n,t)=\mathbbm{1}[a(n,t)\ge0]a(n,t)$ and $a^{-}(n,t)=\mathbbm{1}[a(n,t)<0]|a(n,t)|$. Since $\sum_{n=0}^{\infty}a(n,t)$ is absolutely convergent for any $t$, \[
\sum_{n=0}^{\infty}a(n,t)=\sum_{n=0}^{\infty}a^{+}(n,t)-\sum_{n=0}^{\infty}a^{-}(n,t)
\]
(where both series on the right-hand side converge, hence converge absolutely). We show that $\lim_{t\to0}\sum_{n=0}^{\infty}a^{+}(n,t)=0$, and analogous argument then applies for the series of negative terms.

Let $\varepsilon>0$. First fix some some $t^{*}$. Since $\sum_{n=0}^{\infty}a^{+}(n,t^{*})$ converges absolutely, there exists $\bar{n}$ such that $\sum_{n=\bar{n}+1}^{\infty}a^{+}(n,t^{*})\le\frac{\varepsilon}{2}$. Now fix $\bar{n}$, since $\lim_{t\to0}a^{+}(n,t)=0$ for all $n\in\{0,...,\bar{n}\}$, there exists $\bar{t}\ge t^{*}$ such that $\sum_{n=0}^{\bar{n}}a^{+}(n,t)<\frac{\varepsilon}{2}$ for all $t\ge\bar{t}$. Moreover, since $a(n,t)$ is decreasing in $t$, $\sum_{n=\bar{n}+1}^{\infty}a^{+}(n,t)\le\sum_{n=\bar{n}+1}^{\infty}a^{+}(n,t^{*})=\frac{\varepsilon}{2}$ for all $t\ge t^{*}$. Hence, $\sum_{n=0}^{\infty}a^{+}(n,t)\le\varepsilon$ for all $t\ge\bar{t}$.
\end{proof}

\subsubsection*{Proof of \Cref{Theorem-PopulationUncertainty}}

For notational convenience, let $\mathcal{R}_n(g,m)\equiv \mathcal{M}_n(g,m)Z_n(g,m)\frac{nQ(n)}{\mu}$, where $\mathcal{M}_n(g,m)\equiv \binom{n-1}{m,\tau(n)-g,\tau(n)+g-m}$, and 
\[
Z_n(g,m)=P(G=g|M=m,S_i=s^0,N=n)V_i^d(S_i=s^0,G=g,M=m,N=n),
\]
and let $v^*=\max\{|v^{p^*}-v^{p_*}|:(v^{p^*},v^{p_*})\in \mathcal{V}^{p^*}\times \mathcal{V}^{p_*}\}$.

\begin{proof}
First, suppose $\mathcal{K}_*(n_0)<0$ and fix $\varepsilon\in (0,1)$. We consider two cases.

\medskip

\noindent\emph{Case 1:} Suppose $V^G_0(\tau_0)<0$, which implies $\binom{n_0-1}{\tau_0}Z_0(\tau_0,\tau_0)\frac{n_0Q(n_0)}{\mu}<0$ by \Cref{Assumption-New-RP}. By \Cref{Assumption-RP}, $\lim_{n\to \infty}nQ(n)=0$, and so
\begin{align*}
\sum_{n=n_0+1}^{\infty} \sum_{m=\tau(n)}^{n-1}\binom{n-1}{m} \lambda^{m-\tau_0}(1-\lambda)^{n-1-m-\tau_0}|Z_n(\tau(n),m)|\frac{nQ(n)}{\mu} \le \frac{v^*}{\lambda^{\tau_0}(1-\lambda)^{\tau_0}},
\end{align*}
hence, the series is absolutely convergent. Moreover, for $m\ge \tau(n)$, it follows that $\lambda^{m-\tau_0}(1-\lambda)^{n-1-m-\tau_0}$ is strictly increasing in $\lambda\in (0,1/2)$ and converges to $0$ as $\lambda\to 0$. As a result, there exists $\bar{\lambda}\in (0,1)$ such that, for all $\lambda \in (0,\bar{\lambda})$,
\begin{align*}
\sum_{n=n_0+1}^{\infty} \sum_{m=\tau(n)}^{n-1}\mathcal{R}_n(\tau(n),m) \lambda^{m-\tau_0}(1-\lambda)^{n-1-m-\tau_0} \le \frac{1}{2}\binom{n_0-1}{\tau_0}|Z_0(\tau_0,\tau_0)|\frac{n_0Q(n_0)}{\mu}
\end{align*}
Therefore, for all $\lambda\in (0,\bar{\lambda})$,
\begin{align*}
\Pi(0,\lambda)&=\sum_{n=n_0}^{\infty} \sum_{m=\tau(n)}^{n-1}\mathcal{R}_n(\tau(n),m) \lambda^{m}(1-\lambda)^{n-1-m}\\
&\le \frac{1}{2}\lambda^{\tau_0}(1-\lambda)^{\tau_0}\binom{n_0-1}{\tau_0}Z_0(\tau_0,\tau_0)\frac{n_0Q(n_0)}{\mu} <0.
\end{align*}
Let $\lambda_{\varepsilon}'$ be the unique solution to $\sum_{n=0}^{\infty}(1-\lambda)^nQ(n)=1-\varepsilon$, and $\lambda_{\varepsilon} =\min\{ \bar{\lambda},\lambda_{\varepsilon}'\}$. Then, for all $\lambda \in (0,\lambda_{\varepsilon})$, $\sigma^0$ is an equilibrium in which $\pb $ wins with probability exceeding $1-\varepsilon$. 

\medskip

\noindent\emph{Case 2:} Suppose that $V^G_0(\tau_0)>0$ but $\sum_{\kappa=0}^{\tau_0} \theta^{\kappa}\binom{\tau_0}{\kappa} Z_0(\kappa,\kappa)<0$ for some $\theta\in \mathbbm{R}_{++}$.

For any $\lambda \in (0,\frac{\theta}{1-\theta})$, let $\alpha_{\lambda}\equiv \frac{\lambda}{\theta(1-\lambda)}$; then, $\alpha_{\lambda} \in (0,1)$, is strictly increasing in $\lambda$, and converges to $0$ as $\lambda\to 0$. 

Since $\lim_{n\to \infty}nQ(n)=0$, 
\begin{align*}
\sum_{n=n_0+1}^{\infty} \sum_{g=0}^{\tau(n)} \sum_{m=g}^{g+\tau(n)} \mathcal{R}_n(g,m) \lambda^{m}(1-\lambda)^{n-n_0-m}\alpha_{\lambda}^{\tau(n)-\tau_0-g}(1-\alpha_{\lambda})^{\tau(n)-\tau_0+g-m}\le \frac{v^*}{\alpha_{\lambda}^{\tau_0}(1-\alpha_{\lambda})^{\tau_0}}
\end{align*}
and so the series on the left-hand side is absolutely convergent for any $\lambda \in (0,\frac{\theta}{1+\theta})$. Moreover, 
\begin{align*}
\lambda^{m}(1-\lambda)^{n-n_0-m}&\alpha_{\lambda}^{\tau(n)-\tau_0-g}(1-\alpha_{\lambda})^{\tau(n)-\tau_0+g-m}\\
&= \theta^{-n+n_0+m}(\lambda(1-\lambda)\theta-\lambda^2)^{\tau(n)-\tau_0}\left(\frac{\lambda}{\theta(1-\lambda)-\lambda}\right)^{m-g},
\end{align*}
which is strictly increasing in $\lambda \in (0,\frac{\theta}{2(1+\theta)})$ and converges to $0$ as $\lambda\to 0$. As a result, there exists $\bar{\lambda}\in (0,\frac{\theta}{2(1+\theta)})$ such that, for all $\lambda \in (0,\bar{\lambda})$,
\begin{align*}
\sum_{n=n_0+1}^{\infty} \sum_{g=0}^{\tau(n)} \sum_{m=g}^{g+\tau(n)} &|\mathcal{R}_n(g,m)| \lambda^{m}(1-\lambda)^{n-n_0-m}\alpha_{\lambda}^{\tau(n)-\tau_0-g}(1-\alpha_{\lambda})^{\tau(n)-\tau_0+g-m}\\
&\le \frac{1}{4}\binom{n_0-1}{\tau_0}\frac{n_0Q(n_0)}{\mu}\sum_{\kappa=0}^{\tau_0} \theta^{\kappa}\binom{\tau_0}{\kappa} |Z_0(\kappa,\kappa)|
\end{align*}
Moreover,
\begin{align*}
\sum_{g=0}^{\tau_0} \sum_{m=g+1}^{g+\tau_0} &|\mathcal{R}_{0}(g,m)| \lambda^{m}(1-\lambda)^{-m}\alpha_{\lambda}^{-g}(1-\alpha_{\lambda})^{g-m}\\
&= \sum_{g=0}^{\tau_0}  \sum_{m=g+1}^{g+\tau_0} |\mathcal{R}_{0}(g,m)|\theta^m\left(\frac{\lambda}{\theta(1-\lambda)-\lambda}\right)^{m-g},
\end{align*}
which converges to $0$ as $\lambda\to 0$. Therefore, there exists $\bar{\lambda}'\in (0,\bar{\lambda})$ such that
\begin{align*}
\sum_{g=0}^{\tau_0} \sum_{m=g+1}^{g+\tau_0} &|\mathcal{R}_{0}(g,m)| \lambda^{m}(1-\lambda)^{-m}\alpha_{\lambda}^{-g}(1-\alpha_{\lambda})^{g-m}\\
&\le \frac{1}{4}\binom{n_0-1}{\tau_0}\frac{n_0Q(n_0)}{\mu}\sum_{\kappa=0}^{\tau_0} \theta^{\kappa}\binom{\tau_0}{\kappa} |Z_0(\kappa,\kappa)|
\end{align*}
for all $\lambda \in (0,\bar{\lambda}')$. Therefore, for all $\lambda\in (0,\bar{\lambda}')$,
\begin{align*}
\Pi(\alpha_{\lambda},\lambda)&=\sum_{n=n_0}^{\infty} \sum_{g=0}^{\tau(n)} \sum_{m=g}^{g+\tau(n)} |\mathcal{R}_n(g,m)| \lambda^{m}(1-\lambda)^{n-1-m}\alpha_{\lambda}^{\tau(n)-g}(1-\alpha_{\lambda})^{\tau(n)+g-m}\\
&\le \frac{1}{2}\alpha_{\lambda}^{\tau_0}(1-\alpha_{\lambda})^{\tau_0}(1-\lambda)^{n_0-1}\binom{n_0-1}{\tau_0} \frac{n_0Q(n_0)}{\mu}\sum_{\kappa=0}^{\tau_0} \theta^{\kappa}\binom{\tau_0}{\kappa} Z_0(\kappa,\kappa)<0.
\end{align*}
Finally,  analogous to the argument in Case 1, $V^G_0(\tau_0)>0$ implies that there exists $\bar{\lambda}''\in (0,\bar{\lambda}')$ such that $\Pi(0,\lambda)>0$ for all $\lambda \in (0,\bar{\lambda}'')$. As a result, for any $\lambda \in (0,\bar{\lambda}'')$ there exists $\alpha_{\lambda}'\in (0,\alpha_{\lambda})$ such that $\Pi(\alpha_{\lambda}',\lambda)=0$; hence an equilibrium.

Now let $\lambda_{\varepsilon}'$ be the unique solution to $\sum_{n=n_0}^{\infty}\left(\frac{\theta(1-\lambda)-\lambda}{\theta}\right)^{n}Q(n)=1-\varepsilon$ when $\varepsilon\le \theta^{-1}$ and $1$ otherwise, and let $\lambda_{\varepsilon}=\min\{\bar{\lambda}'',\lambda_{\varepsilon}'\}$. Then, for all $\lambda \in (0,\lambda_{\varepsilon})$, $\sigma^{\alpha_{\lambda}'}$ is an equilibrium in which $\pb $ wins with probability exceeding $1-\varepsilon$. 

\medskip

\noindent Now, suppose $\mathcal{K}_*(n_0)>0$ and fix $\varepsilon\in (0,1)$. The advantageous correlation condition implies that $Z_0(\tau_0,\tau_0),Z_0(0,0)>0$, and therefore there exists $\delta>0$ such that $\sum_{\kappa=0}^{\tau_0}\theta^{\kappa}\binom{\tau_0}{\kappa}Z_0(\kappa,\kappa)>\delta$ for all $\theta \in \mathbbm{R}_{++}$.

Let $\nu_{\varepsilon}$ be the unique solution in $(0,1)$ to $\sum_{n=n_0}^{\infty}\nu^nQ(n)=1-\varepsilon$, and let $\bar{\lambda}=1-\sqrt{\nu_{\varepsilon}}$ and $\bar{\alpha}=\sqrt{\nu_{\varepsilon}}$. Then, for any $\alpha \in (\bar{\alpha},1]$ and $\lambda\in (0,\bar{\lambda})$, $\pg $ wins with probability exceeding 
\[
\sum_{n=n_0}^{\infty}\bar{\alpha}^n(1-\bar{\lambda})^n Q(n)= \sum_{n=n_0}^{\infty}\nu_{\varepsilon}^nQ(n)=1-\varepsilon
\]
in the strategy profile $\sigma^{\alpha}$. It therefore suffices to show that there exists $\lambda_{\varepsilon}\in (0,\bar{\lambda})$ such that, for all $\lambda \in (0,\lambda_{\varepsilon})$ and $\alpha \in [0,\bar{\alpha}]$, $\sigma^{\alpha}$ is not an equilibrium. We do this by first showing that there exists $\bar{\lambda}_0 \in (0,1)$ such that $\sigma^0$ is not an equilibrium for all $\lambda \in (0,\bar{\lambda}_0)$ (step 1), and then showing that there exists $\lambda^*\in (0,1)$ such that, for all $\lambda\in (0,\lambda^*)$, $\sigma^{\alpha}$ is not an equilibrium for any $\alpha \in (0,\bar{\alpha})$ (step 2). 

\medskip

\noindent \emph{Step 1:} Since $\lim_{n\to \infty}nQ(n)=0$, 
\begin{align*}
\sum_{n=n_0+1}^{\infty}\sum_{m=\tau(n)}^{n-1}|\mathcal{R}_n(\tau(n),m)|\binom{n-1}{m}\lambda^{m-\tau_0}(1-\lambda)^{n-1-m-\tau_0}\le \frac{v^*}{\lambda^{\tau_0}(1-\lambda)^{\tau_0}}
\end{align*}
and so the series is absolutely convergent. Moreover, for $m\ge \tau(n)>\tau_0$, $\lambda^{m-\tau_0}(1-\lambda)^{n-1-m-\tau_0}$ is strictly increasing in $\lambda \in (0,1/2)$ and converges to $0$ as $\lambda\to 0$. As a result, there exists $\lambda_0 \in (0,1)$ such that 
\begin{align*}
\sum_{n=n_0+1}^{\infty}\sum_{m=\tau(n)}^{n-1}&|\mathcal{R}_{n}(\tau(n),m)|\lambda^{m-\tau_0}(1-\lambda)^{n-1-m-\tau_0}\le \frac{1}{4}\binom{n_0-1}{\tau_0}Z_0(\tau_0,\tau_0)\frac{n_0Q(n_0)}{\mu}
\end{align*}
Moreover, since $\lambda^{m-\tau_0}(1-\lambda)^{\tau_0-m}$ is increasing in $\lambda\in (0,1)$ and converges to $0$ as $\lambda\to 0$, there exists $\lambda_0'\in (0,1)$ such that 
\begin{align*}
\sum_{m=\tau_0+1}^{n_0-1} \binom{n_0-1}{m} \lambda^{m-\tau_0}(1-\lambda)^{\tau_0-m} |Z_0(\tau_0,m)|\le \frac{1}{4}\binom{n_0-1}{\tau_0}Z_0(\tau_0,\tau_0)
\end{align*}
Let $\bar{\lambda}_0 =\min \{\lambda_0,\lambda_0'\}$; then for all $\lambda \in (0,\bar{\lambda}_0)$, 
\begin{align*}
\Pi(0,\lambda)&=\sum_{n=n_0}\sum_{m=\tau(n)}\mathcal{R}_{n}(\tau(n),m)\binom{n-1}{m}\lambda^{m}(1-\lambda)^{n-1-m}\\
&\ge  \frac{1}{2}\binom{n_0-1}{\tau_0}\lambda^{\tau_0}(1-\lambda)^{\tau_0}Z_0(\tau_0,\tau_0)\frac{n_0Q(n_0)}{\mu}>0,
\end{align*}
and so $\sigma^0$ is not an equilibrium.

\medskip

\noindent \emph{Step 2:} It remains to show that there exists $\lambda^*\in (0,1)$ such that, for all $\lambda \in (0,\lambda^*)$, $\sigma^{\alpha}$ is not an equilibrium for any $\alpha \in (0,\bar{\alpha})$. We show this by establishing a contradiction. Suppose that, for any $\lambda^*\in (0,1)$, there exists $\lambda \in (0,\lambda^*)$ and $\alpha_{\lambda}\in (0,\bar{\alpha})$ such that $\Pi(\alpha_{\lambda},\lambda)=0$; hence, there exists a sequence $(\lambda_t,\alpha_t)_{t=1}^{\infty}$ such that $\lambda_t\to 0$ and, for all $t\ge1$, $\alpha_t \in (0,\bar{\alpha})$, and $\Pi(\alpha_t,\lambda_t)=0$, where
\begin{align}\label{Equation_PUC}
\Pi(\alpha_t,\lambda_t)=\sum_{n=n_0}^{\infty} \sum_{g=0}^{\tau(n)} \sum_{m=g}^{g+\tau(n)} \mathcal{R}_n(g,m) \lambda_t^{m}(1-\lambda_t)^{n-1-m}\alpha_t^{\tau(n)-g}(1-\alpha_t)^{\tau(n)+g-m}.
\end{align}

We consider three collectively exhaustive cases: (i) there is a subsequence such that $\frac{\alpha_t(1-\lambda_t)}{\lambda_t}\to 0$, (ii) there is a subsequence such that $\alpha_t\to 0$ but $\frac{\alpha_t(1-\lambda_t)}{\lambda_t}\ge  \gamma$ for some $\gamma>0$, and (iii) there is a subsequence such that $\alpha_t \ge \gamma$ for some $\gamma>0$.

\medskip

\noindent \emph{Case (i).} In this case, there is a subsequence such that  $\lambda_{t},\alpha_{t},\frac{\alpha_{t}(1-\lambda_{t})}{\lambda_{t}},\lambda_{t}(1-\lambda_{t})(1-\alpha_{t}),\frac{\lambda_{t}}{(1-\lambda_{t})(1-\alpha_{t})}$ are all decreasing, and converge to $0$. From $\Pi(\alpha_t,\lambda_t)=0$, it follows that (for all $t$, with the subscript suppressed for convenience),
\begin{align*}
-&\binom{n_{0}-1}{\tau_{0}}\lambda^{\tau_{0}}(1-\lambda)^{\tau_{0}}(1-\alpha)^{\tau_{0}}Z_{0}(\tau_{0},\tau_{0})\frac{n_0Q(n_{0})}{\mu}\\
& =\sum_{g=0}^{\tau_{0}-1}\mathcal{R}_{0}(g,g)\lambda^{g}(1-\lambda)^{n_{0}-1-g}\alpha^{\tau_{0}-g}(1-\alpha)^{\tau_{0}}\\
 & +\sum_{g=0}^{\tau_{0}}\sum_{m=g+1}^{g+\tau_{0}}\mathcal{R}_{0}(g,m)\lambda^{m}(1-\lambda)^{n_{0}-1-m}\alpha^{\tau_{0}-g}(1-\alpha)^{\tau_{0}+g-m}\\
 & +\sum_{n=n_{0}+1}^{\infty}\mathcal{R}_{n}(\tau(n),\tau(n))\lambda^{\tau(n)}(1-\lambda)^{\tau(n)}(1-\alpha)^{\tau(n)}\\
 & +\sum_{n=n_{0}+1}^{\infty}\sum_{g=0}^{\tau(n)-1}\mathcal{R}_{n}(g,g)\lambda^{g}(1-\lambda)^{n-1-g}\alpha^{\tau(n)-g}(1-\alpha)^{\tau(n)}\\
 & +\sum_{n=n_{0}+1}^{\infty}\sum_{g=0}^{\tau(n)}\sum_{m=g+1}^{g+\tau(n)}\mathcal{R}_{n}(g,m)\lambda^{m}(1-\lambda)^{n-1-m}\alpha^{\tau(n)-g}(1-\alpha)^{\tau(n)+g-m}
\end{align*}
Therefore (dividing both sides by $[\lambda(1-\lambda)(1-\alpha)]^{\tau_{0}}$),
\begin{align*}
-&\binom{n_{0}-1}{\tau_{0}}Z_{0}(\tau_{0},\tau_{0})\frac{n_0Q(n_{0})}{\mu}\\
 & =\sum_{g=0}^{\tau_{0}-1}\mathcal{R}_{0}(g,g)\left(\frac{\alpha(1-\lambda)}{\lambda}\right)^{\tau_{0}-g}\\
 & +\sum_{g=0}^{\tau_{0}}\sum_{m=g+1}^{g+\tau_{0}}\mathcal{R}_{0}(g,m)\left(\frac{\alpha(1-\lambda)}{\lambda}\right)^{\tau_{0}-g}\left(\frac{\lambda}{(1-\lambda)(1-\alpha)}\right)^{m-g}\\
 & +\sum_{n=n_{0}+1}^{\infty}\mathcal{R}_{n}(\tau(n),\tau(n))[\lambda(1-\alpha)(1-\lambda)]^{\tau(n)-\tau_{0}}\\
 & +\sum_{n=n_{0}+1}^{\infty}\sum_{g=0}^{\tau(n)-1}\mathcal{R}_{n}(g,g)\left(\frac{\alpha(1-\lambda)}{\lambda}\right)^{\tau(n)-g}[\lambda(1-\alpha)(1-\lambda)]^{\tau(n)-\tau_{0}}\\
 & +\sum_{n=n_{0}+1}^{\infty}\sum_{g=0}^{\tau(n)}\sum_{m=g+1}^{g+\tau(n)}\mathcal{R}_{n}(g,m)\left(\frac{\alpha(1-\lambda)}{\lambda}\right)^{\tau(n)-g}\left(\frac{\lambda}{(1-\lambda)(1-\alpha)}\right)^{m-g}[\lambda(1-\alpha)(1-\lambda)]^{\tau(n)-\tau_{0}}.
\end{align*}
By \Cref{Lemma-S}, the left-hand side converges to $0$ but the right-hand side is constant and bounded away from $0$. 
\medskip

\noindent \emph{Case (ii).}  In this case, there is a subsequence such that $\lambda_{t},\alpha_{t},\frac{\alpha_t}{1-\alpha_t},\alpha_t(1-\lambda_t)^{2}(1-\alpha_t)$ are all decreasing and converge to $0$. From $\Pi(\alpha_t,\lambda_t)=0$ it follows that ($t$ subscript suppressed)
\begin{align*}
-&\sum_{g=0}^{\tau_{0}}\binom{n_{0}-1}{g}\binom{n_{0}-1-g}{\tau_{0}-g}\lambda^{g}(1-\lambda)^{n_{0}-1-g}\alpha^{\tau_{0}-g}(1-\alpha)^{\tau_{0}}Z_{0}(g,g)\frac{n_0Q(n_{0})}{\mu}\\
= & \sum_{g=0}^{\tau_{0}}\sum_{m=g+1}^{\tau_{0}}\mathcal{R}_{0}(g,m)\lambda^{m}(1-\lambda)^{n_{0}-1-m}\alpha^{\tau_{0}-g}(1-\alpha)^{\tau_{0}+g-m}\\
+ & \sum_{n=n_{0}+1}^{\infty}\sum_{g=0}^{\tau(n)}\mathcal{R}_{n}(g,g)\lambda^{g}(1-\lambda)^{n-1-g}\alpha^{\tau(n)-g}(1-\alpha)^{\tau(n)}\\
+ & \sum_{n=n_{0}+1}^{\infty}\sum_{g=0}^{\tau(n)}\sum_{m=g+1}^{g+\tau(n)}\mathcal{R}_{n}(g,m)\lambda^{m}(1-\lambda)^{n-1-m}\alpha^{\tau(n)-g}(1-\alpha)^{\tau(n)+g-m}
\end{align*}
Therefore (dividing both sides by $\alpha^{\tau_{0}}(1-\alpha)^{\tau_{0}}(1-\lambda)^{n_{0}-1}$),
\begin{align*}
 -& \binom{n_{0}-1}{\tau_{0}}\frac{n_0Q(n_{0})}{\mu}\sum_{g=0}^{\tau_{0}}\binom{\tau_{0}}{g}\left(\frac{\lambda}{\alpha(1-\lambda)}\right)^{g}Z_{0}(g,g)\\
&= \sum_{g=0}^{\tau_{0}}\sum_{m=g+1}^{\tau_{0}}\mathcal{R}_{0}(g,m)\left(\frac{\lambda}{\alpha(1-\lambda)}\right)^{m}\left(\frac{\alpha}{1-\alpha}\right)^{m-g}\\
 &+ \sum_{n=n_{0}+1}^{\infty}\sum_{g=0}^{\tau(n)}\mathcal{R}_{n}(g,g)\left(\frac{\lambda}{\alpha(1-\lambda)}\right)^{g}[\alpha(1-\alpha)(1-\lambda)^2]^{\tau(n)-\tau_{0}}\\
&+ \sum_{n=n_{0}+1}^{\infty}\sum_{g=0}^{\tau(n)}\sum_{m=g+1}^{g+\tau(n)}\mathcal{R}_{n}(g,m)\left(\frac{\lambda}{\alpha(1-\lambda)}\right)^{m}\left(\frac{\alpha}{1-\alpha}\right)^{m-g}[\alpha(1-\alpha)(1-\lambda)^2]^{\tau(n)-\tau_{0}}\\
\equiv& \tilde{\Pi}(\alpha,\lambda)
\end{align*}

If there exists a further subsequence such that $\frac{\lambda_t}{\alpha_t(1-\lambda_t)}$ is decreasing, then the left-hand side converges to $0$ by \Cref{Lemma-S} while the right-hand side is constant. Otherwise, there exists a subsequence such that $\frac{\lambda_t}{\alpha_t(1-\lambda_t)}$ converges up to some $\theta^*>0$. For each $t$ in that subsequence, let $\alpha^*_t = \frac{\lambda_t}{\theta^*(1-\lambda_t)}$.  Eventually, $\alpha^*_t\in (0,1)$, and so 
\begin{align*}
  & \sum_{g=0}^{\tau_{0}}\sum_{m=g+1}^{\tau_{0}}\mathcal{R}_{0}(g,m)\left(\theta^*\right)^{m}\left(\frac{\alpha^*}{1-\alpha^*}\right)^{m-g}\\
 +& \sum_{n=n_{0}+1}^{\infty}\sum_{g=0}^{\tau(n)}\mathcal{R}_{n}(g,g)\left(\theta^*\right)^{g}[\alpha^*(1-\alpha^*)(1-\lambda)^2]^{\tau(n)-\tau_{0}}\\
+& \sum_{n=n_{0}+1}^{\infty}\sum_{g=0}^{\tau(n)}\sum_{m=g+1}^{g+\tau(n)}\mathcal{R}_{n}(g,m)\left(\theta^*\right)^{m}\left(\frac{\alpha^*}{1-\alpha^*}\right)^{m-g}[\alpha^*(1-\alpha^*)(1-\lambda)^2]^{\tau(n)-\tau_{0}}
\end{align*}
is  absolutely convergent. Since, for each $t$ there exists $t'\ge t$ such that $\frac{\alpha_{t'}}{1-\alpha_{t'}}\le \frac{\alpha_{t}}{1-\alpha_{t}}$ and $\alpha_{t'}(1-\alpha_{t'})(1-\lambda_{t'})^2\le \alpha^*_t(1-\alpha^*_t)(1-\lambda_t)^2$, it follows that 
\begin{align*}
  & \sum_{g=0}^{\tau_{0}}\sum_{m=g+1}^{\tau_{0}}\mathcal{R}_{0}(g,m)\left(\theta^*\right)^{m}\left(\frac{\alpha}{1-\alpha}\right)^{m-g}\\
 +& \sum_{n=n_{0}+1}^{\infty}\sum_{g=0}^{\tau(n)}\mathcal{R}_{n}(g,g)\left(\theta^*\right)^{g}[\alpha(1-\alpha)(1-\lambda)^2]^{\tau(n)-\tau_{0}}\\
+& \sum_{n=n_{0}+1}^{\infty}\sum_{g=0}^{\tau(n)}\sum_{m=g+1}^{g+\tau(n)}\mathcal{R}_{n}(g,m)\left(\theta^*\right)^{m}\left(\frac{\alpha}{1-\alpha}\right)^{m-g}[\alpha(1-\alpha)(1-\lambda)^2]^{\tau(n)-\tau_{0}}
\end{align*}
is eventually absolutely convergent, and then converges $0$ by \Cref{Lemma-S}.

\medskip

\noindent \emph{Case (iii).} In this case, there is a subsequence such that $\lambda_{t}$ is decreasing and, since $\alpha \in (0,\bar{\alpha})$, there exists some $\gamma\in (0,1/2)$ such that $\alpha_t\in[\gamma,(1-\gamma)]$ for all $t$. From $\Pi(\alpha_t,\lambda_t)=0$ it follows that ($t$ subscript suppressed)
\begin{align*}
 -& \sum_{n=n_{0}}^{\infty}\binom{n-1}{\tau(n)}(1-\lambda)^{n-1}\alpha^{\tau(n)}(1-\alpha)^{\tau(n)}Z_{n}(0,0)\frac{nQ(n)}{\mu}\\
&= \sum_{n=n_{0}}^{\infty}\sum_{m=1}^{\tau(n)}\mathcal{R}_{n}(0,m)\lambda^{m}(1-\lambda)^{n-1-m}\alpha^{\tau(n)}(1-\alpha)^{\tau(n)-m}\\
&+ \sum_{n=n_{0}}^{\infty}\sum_{g=1}^{\tau(n)}\sum_{m=g}^{g+\tau(n)}\mathcal{R}_{n}(g,m)\lambda^{m}(1-\lambda)^{n-1-m}\alpha^{\tau(n)-g}(1-\alpha)^{\tau(n)+g-m}
\end{align*}
Since $\alpha\in[\delta,1-\delta]$ it follow that $\alpha^{\tau(n)}(1-\alpha)^{\tau(n)}\ge\gamma^{\tau(n)}$, and so the left-hand side is greater $\sum_{n=n_{0}}^{\infty}\binom{n-1}{\tau(n)}(1-\lambda)^{n-1}\delta^{\tau(n)}Z_{n}(0,0)\frac{nQ(n)}{\mu}$, which converges to 
\[
\sum_{n=n_{0}}^{\infty}\binom{n-1}{\tau(n)}\delta^{\tau(n)}Z_{n}(0,0)Q(n)>0,
\]
 while the right-hand side converges to $0$ by \Cref{Lemma-S}.
 \end{proof}

\subsubsection{The Role of Elites (\Cref{Section-Elite})}

\paragraph{Preliminaries:} For any state $\omega$,  we denote by $G_{E}(\omega)$ the number of elites who receive good news,  $M_{E}(\omega)$ the number of elites who receive informative signals, $G_{N}(\omega)=G(\omega)-G_E(\omega)$ and $M_{N}(\omega)=M(\omega)-M_E(\omega)$, with typical realizations of these random variables denoted, respectively, by $g_E$, $m_E$, $g_{N}$,  and $m_{N}$. 

For $g_E \in \{0,..., |\mathcal{E}|\}$, $g_{N}\in \{0,...,|\mathcal{NE}|\}$, $m_E \in \{g_e,...,|\mathcal{E}|\}$, and $m_{N}\in \{g_e,...,|\mathcal{NE}|\}$,
\begin{align*}
Z_i(g_E,g_{N},m_E,m_{N})\equiv P(g_E,g_{N}|S_i=s^0,m_E,m_{N})V_i(S_i=s^0,g_E,g_{N},m_E,m_{N}).
\end{align*}

\subsubsection*{Proof of \Cref{Prop-MajNegcor}}

\begin{proof}[\unskip\nopunct]
Fix $\varepsilon \in (0,1)$ and let $\sigma^* \in \Sigma^*$ with $\sigma^*_i(s^0)=\mathbbm{1}[i\in\mathcal{E}]$. Since $|\mathcal{E}|\le \tau$, $\pb $ wins for the strategy profile $\sigma^*$ in the event $\{S=s_0\}$, and therefore wins with probability exceeding $1-\varepsilon$ for all $\lambda \in (0, 1-(1-\varepsilon)^{\frac{1}{n}})$. Hence, it is sufficient to show that $\sigma^*$ is a strict equilibrium for $\lambda$ sufficiently small. 


If $i \in \mathcal{E}$ receives signal $s^0$, then
\begin{align*}
\Pi_i(\sigma^*,\lambda) & =\lambda^{\tau-|\mathcal{E}|+1}\sum_{g_{E}=0}^{|\mathcal{E}|-1}\sum_{m_E=g_E}^{|\mathcal{E}|-1} \sum_{m_N=\hat{g}(m_E,m_N)}^{|\mathcal{NE}|} \binom{|\mathcal{E}|-1}{m_E}\binom{|\mathcal{NE}|}{m_{N}}\\
&\hspace{2.7cm} \lambda^{m_E+m_N-\tau+|\mathcal{E}|-1}(1-\lambda)^{n-1-m_E-m_N}Z_i(g_E,\hat{g}(m_E,m_N),m_E,m_{N}),
\end{align*}
where $\hat{g}(m_E,m_N)=\tau-(|\mathcal{E}|-1-(m_e-g_e))$.  Since
\begin{align*}
\lim_{\lambda\to 0} \; \Pi_i(\sigma^*,\lambda)\lambda^{-\tau+|\mathcal{E}|-1}=\binom{|\mathcal{NE}|}{\tau-|\mathcal{E}|+1}Z_i(0,\tau-|\mathcal{E}|+1,0,\tau-|\mathcal{E}|+1),
\end{align*}
which is strictly positive by elite-adverse correlation, there exists $\bar{\lambda}_E\in (0,1)$ such that $\Pi_i(\sigma^*,\lambda)>0$ for all elites who receive the signal $s^0$ for all $\lambda \in (0,\bar{\lambda}_E)$.

If $i \in \mathcal{NE}$ receives signal $s^0$, then
\begin{align*}
\Pi_i(\sigma^*,\lambda) & =\lambda^{\tau-|\mathcal{E}|}\sum_{g_{E}=0}^{|\mathcal{E}|}\sum_{m_E=g_E}^{|\mathcal{E}|} \sum_{m_N=\hat{g}(m_E,m_N)-1}^{|\mathcal{NE}|-1} \binom{|\mathcal{E}|}{m_E}\binom{|\mathcal{NE}|-1}{m_{N}}\\
&\hspace{2.2cm} \lambda^{m_E+m_N-\tau+|\mathcal{E}|}(1-\lambda)^{n-1-m_E-m_N}Z_i(g_E,\hat{g}(m_E,m_N)-1,m_E,m_{N}).
\end{align*}
Since
\begin{align*}
\lim_{\lambda\to 0} \; \Pi_i(\sigma^*,\lambda)\lambda^{-\tau+|\mathcal{E}|}=\binom{|\mathcal{NE}|}{\tau-|\mathcal{E}|}Z_i(0,\tau-|\mathcal{E}|,0,\tau-|\mathcal{E}|),
\end{align*}
which is strictly negative by elite-adverse correlation, there exists $\bar{\lambda}_{NE}\in (0,1)$ such that $\Pi_i(\sigma^*,\lambda)<0$ for all non-elites who receive the signal $s^0$ for all $\lambda \in (0,\bar{\lambda}_{NE})$. 

As a result, $\sigma^*$ is an equilibrium for all $\lambda\in (0, \min\{\bar{\lambda}_E,\bar{\lambda}_{NE}\})$.
\end{proof}

\subsubsection*{Proof of \Cref{Prop-EliteComparative}}

\begin{proof}[\unskip\nopunct]
For some $(P_W,v_W,v_L,e)$, let $i\in \mathcal{NE}$ and $w\in \{\tau+1,...,n\}$. Then,
\begin{align*}
P(G_N&=\tau-e|S_i=s^0,M=M_N=\tau-e,W=w)\\
&=\left(\frac{w-e}{n-e}\right) \frac{\binom{w-e-1}{\tau-e}\binom{n-1-e-(w-e-1)}{0}}{\binom{n-1-e}{\tau-e} }+\left(\frac{n-w}{n-e}\right) \frac{\binom{w-e}{\tau-e}\binom{n-1-e-(w-e)}{0}}{\binom{n-e-1}{\tau-e}} = \frac{\binom{w-e}{\tau-e}}{\binom{n-e}{\tau-e}}\\
P(G_N&=\tau-e|S_i=s^0,M=M_N=\tau-e,W=\tau)=\frac{\binom{\tau-e}{\tau-e}}{\binom{n-e}{\tau-e}}
\end{align*}
Therefore, for any $w\in \{\tau,...,n\}$,
\begin{align*}
P(W=w|E_0(e))& = \frac{\binom{w-e}{\tau-e} P_W(w)}{ \sum_{w'=\tau}^n \binom{w'-e}{\tau-e} P_W(w')} .
\end{align*}
where $E_0(e)=\{S_i=s^0,G=M=M_N=\tau-e\}$ for $i\in \mathcal{NE}$. Since
 \begin{align*}
 P(W_i|E_0(e),W=w)=\frac{w-e-(\tau-e)}{n-e-(\tau-e)}=\frac{w-\tau}{n-\tau},
 \end{align*}
it follows that 
\begin{align*}
V_i(E_0(e))= \sum_{w=\tau}^n \left(\left(\frac{w-\tau}{n-\tau}\right) v_W - \left(\frac{n-w}{n-\tau}\right)v_L \right)\frac{\binom{w-e}{\tau-e} P_W(w)}{ \sum_{w'=\tau}^n \binom{w'-e}{\tau-e} P_W(w')}.
\end{align*}
For $0\le e < e' \le \tau$ and $\tau \le w < w' \le n$,
\begin{align*}
\binom{w'-e'}{\tau-e'}\binom{w-e}{\tau-e}< \binom{w-e'}{\tau-e'}\binom{w'-e}{\tau-e}
\end{align*}
and so $P_W(.|E_0(e)) \succeq_{LR} P_W(.|E_0(e'))$. Since $P(W_i|E_0(e),W=w)$ is strictly increasing in $w$ for $i\in \mathcal{NE}$, it follows that $V_i(E_0(e)) \ge V_i(E_0(e'))$.
\end{proof}


\end{document}